\documentclass[]{aa}
\usepackage{graphicx}
\usepackage[switch]{lineno}
\usepackage{xcolor}
\usepackage{txfonts}
\usepackage[colorlinks=true, linkcolor=blue, citecolor=blue, urlcolor=blue]{hyperref}

\begin{document} 

   \title{The stellar-to-halo mass relation of central galaxies across three orders of halo mass}

   \author{Victoria Toptun\inst{1}\thanks{victoria.toptun@eso.org}
   \and Paola Popesso \inst{1,}\inst{2}
    \and Ilaria Marini\inst{2,}\inst{1}
    \and Stephan Vladutescu-Zopp\inst{1}
    \and Klaus Dolag \inst{3,}\inst{4,}\inst{2}
    \and Peter Behroozi \inst{5}
    \and Lorenzo Lovisari\inst{6,}\inst{7,}
    \and Stefano Ettori\inst{8,}\inst{9}
    \and Veronica Biffi\inst{10,}\inst{11}
    \and Xiaohu Yang\inst{12,13}
    \and Natanael de Isídio\inst{1}
    \and Daudi T. Mazengo\inst{1,}\inst{14}}
    
   \institute{European Southern Observatory, Karl Schwarzschildstrasse 2, 85748, Garching bei M\"unchen, Germany
    \and Excellence Cluster ORIGINS, Boltzmannstr. 2, D-85748 Garching bei M\"unchen, Germany
    \and Universitäts-Sternwarte, Fakultät für Physik, Ludwig-Maximilians-Universität München, Scheinerstr.1, 81679 München, Germany
    \and Max-Planck-Institut für Astrophysik, Karl-Schwarzschildstr. 1, 85741 Garching bei M\"unchen, Germany
    \and Department of Astronomy and Steward Observatory, University of Arizona, Tucson, AZ 85721, USA
    \and INAF, Istituto di Astrofisica Spaziale e Fisica Cosmica di Milano, via A. Corti 12, 20133 Milano, Italy
    \and Center for Astrophysics $|$ Harvard $\&$ Smithsonian, 60 Garden Street, Cambridge, MA 02138, USA
    \and INAF, Osservatorio di Astrofisica e Scienza dello Spazio, via Piero Gobetti 93/3, 40129 Bologna, Italy
    \and INFN, Sezione di Bologna, viale Berti Pichat 6/2, 40127 Bologna, Italy 
    \and  INAF, Osservatorio Astronomico di Trieste, Via Tiepolo 11, 34143 Trieste, Italy
    \and IFPU, Institute for Fundamental Physics of the Universe, Via Beirut 2, I-34014 Trieste, Italy
    \and State Key Laboratory of Dark Matter Physics, Tsung-Dao Lee Institute \& School of Physics and Astronomy, Shanghai Jiao Tong University, Shanghai 201210, China
    \and Shanghai Key Laboratory for Particle Physics and Cosmology, and Key Laboratory for Particle Physics, Astrophysics and Cosmology, Ministry of Education, Shanghai Jiao Tong University, Shanghai 200240, China
    \and Physics Department, College of Natural and Mathematical Sciences, P.O.Box 338, The University of Dodoma, Tanzania}

   \authorrunning{Toptun et al.}
   \date{Received ; accepted }
 
  \abstract
  {The stellar content of galaxies is tightly connected to the mass and growth of their host dark matter halos. Observational constraints on this relation remain limited, particularly for low-mass groups, leaving uncertainties in how galaxies assemble their stars across halo mass scales. Accurately measuring the {brightest central galaxy (BCG)} stellar-to-halo mass relation (SHMR) over a wide mass range is therefore crucial for understanding galaxy formation and the role of feedback processes. Here we present the SHMR spanning $M_{\rm halo} \sim 10^{12}$--$10^{15}\,M_\odot$, using halo masses derived from eROSITA eRASS1 X-ray data and BCG stellar masses based on SDSS photometry. By stacking X-ray spectra of optically selected groups, we {obtain} robust average halo gas temperatures for each bin, which are then converted to halo masses via the \mbox{$M$--{$T_\mathrm{X}$}} relation. We find that the SHMR peaks near $M_{\rm halo} \sim 10^{12}\,M_\odot$, with a declining stellar fraction at higher masses. {This trend reflects a combination of processes that reduce the efficiency of stellar mass growth in massive halos, such as active galactic nucleus feedback, reduced cooling efficiency, and the increasing dominance of ex situ assembly}, while halos continue to grow through mergers and accretion. Our measurements are consistent {over the full mass range} with previous observational studies, including weak lensing, X-ray analyses of individual clusters, and kinematical and dynamical methods. Comparisons with hydrodynamical simulations show good agreement at low masses but reveal significant discrepancies {in the normalization} at cluster scales, highlighting the sensitivity of BCG stellar growth to feedback prescriptions and halo assembly history. These results provide the first X-ray-based observational SHMR covering three orders of magnitude in halo mass and establish a robust benchmark for testing galaxy formation models.
}


   \maketitle

\section{Introduction\label{sec:intro}}

Galaxy formation is fundamentally linked to the growth of dark matter halos, creating a strong correlation between galaxy properties, especially stellar mass, and the mass and gravitational potential of their host halos \citep{2009ApJ...696..620C, 2010ApJ...717..379B, 2010ApJ...710..903M}. Central galaxies, which reside at the centers of these halos, are particularly shaped by the mass and structure of their hosts. While most galaxies live in relatively small halos, within the galaxy group regime, which contain the majority of the Universe’s baryonic matter {\citep[98\% of systems;][]{yang_galaxy_2007, robotham,paola_24}}, the majority of studies on halo-driven galaxy evolution have focused on the small minority that reside in massive galaxy clusters {above $M_{\mathrm{halo}}\sim 10^{14}\,M_\odot$}. In this work, we extend the scope of such studies for the first time across three orders of magnitude in halo mass, targeting group-scale systems down to halos comparable in mass to that of the Milky Way.

A central example of the galaxy--halo connection is the stellar-to-halo mass relation (SHMR), which quantifies how the stellar mass of central galaxies scales with the total mass of their dark matter halos. Semiempirical models show that the stellar-to-halo mass ratio for central galaxies increases steeply from low-mass halos, which host dwarf systems, up to a peak near $M_{500} \sim 10^{12}$--$10^{12.5}\,M_{\odot}$, characteristic of halos similar in mass to the Local Group. Beyond this scale, the ratio declines with increasing halo mass, indicating that the efficiency of converting baryons into stars drops sharply in more massive halos and becomes very low in systems such as {massive} galaxy clusters \citep{2013MNRAS.428.3121M, 2018MNRAS.477.1822M, 2015MNRAS.450.1604L, 2017MNRAS.470..651R, 2013ApJ...777L..10B, 2014ApJ...793...12B, 2019MNRAS.488.3143B}.

This characteristic shape of the SHMR {captures how stellar mass assembly in central galaxies is regulated by the shifting balance of feedback processes across halo mass scales.} In low-mass halos {below $M_{\mathrm{halo}}\sim 10^{12}\,M_\odot$}, typically hosting a single central galaxy, supernova-driven winds and stellar feedback dominate but lose effectiveness as halo mass increases and the gravitational potential deepens. At this stage, gas accretion is still insufficient to trigger efficient active galactic nucleus (AGN) activity \citep{2024ApJ...971...69D, 2025A&A...699A.311M}.  In higher-mass halos {above $M_{\mathrm{halo}}\sim 10^{13}\,M_\odot$}, AGN feedback becomes the dominant quenching mechanism, suppressing further star formation \citep{2018AstL...44....8K, 2019A&A...631A.175E, 2022ApJ...928...28G}, {with long gas cooling times further limiting the supply of cold gas. Around the peak mass, where stellar and AGN feedback are comparably effective, halos reach their maximum efficiency in converting baryons into stars}. 

{The AGN feedback is likely not the only process shaping this trend. Decreasing gas cooling efficiency in massive halos, the increasing importance of ex situ (merger-driven) stellar mass assembly, and halo assembly bias can all contribute to the observed behavior of the SHMR.} {The presence of the SHMR peak cannot be explained solely by AGN-driven gas depletion and the associated decrease in $M_{\mathrm{halo}}$: gas fractions in systems with $M_{\mathrm{halo}}\sim 10^{13}\,M_\odot$ are typically $\sim3\%$ \citep{2009A&A...498..361P,2016A&A...592A..12E,paola_gasfraction} and can reach the universal baryon fraction ($\Omega_\mathrm{b}/\Omega_\mathrm{m} = 0.154$, \citealt{2020A&A...641A...6P})
 in massive clusters, which is insufficient to account for the more than order-of-magnitude variation observed in the SHMR.}  {Moreover, the} shape of the SHMR reflects the {cumulative} influence {on the brightest central galaxy (BCG) across the history of the system, rather than the instantaneous state of the present epoch}. This is the reason why the SHMR is considered a fundamental relation in galaxy evolution.

Nevertheless, the SHMR remains only partially constrained, primarily due to the difficulty of accurately measuring halo masses. While stellar masses are available for large galaxy samples in surveys such as SDSS, DESI, and GAMA, precise and consistent halo mass estimates are typically limited to the most massive systems, galaxy clusters, where reliable mass proxies exist, {but such systems represent only a small fraction of the population.}
 
Weak gravitational lensing (WL) directly probes the total projected mass distribution through the distortion of background galaxies. It has been widely applied to constrain the SHMR over a broad range of halo masses, from galaxies to massive clusters \citep{2006MNRAS.368..715M,2016MNRAS.457.3200M,2005ApJ...635...73H,2012MNRAS.425.2610R,2015MNRAS.447..298H,2011A&A...534A..14V,2016MNRAS.459.3251V,2020A&A...642A..83D, Wang_2022}. The primary advantage of WL is that it is independent of baryonic tracers, although it requires deep, wide-field imaging and careful control of systematic effects, including shear calibration and accurate source redshift estimation.

Indirect halo mass estimates rely on observables correlated with halo mass, such as the richness, magnitude gap, or velocity dispersion, calibrated against external measurements \citep{2019ApJ...878...14G, 2022ApJ...928...28G,2009ApJ...699.1333H,2011MNRAS.410..210M,2014A&A...561A..79V,2018ApJ...860....2G}. These methods cover groups to massive clusters ($M_{\rm halo} \sim 10^{13}$--$10^{15} M_\odot$). However, they have a large intrinsic scatter and require external calibration, making them indirectly dependent on stellar mass.

Semi-analytical approaches, such as abundance matching and empirical modeling, connect galaxies to halos statistically by matching observed stellar mass or luminosity functions to simulated halo mass functions \citep{Yang_2009, 2009ApJ...696..620C, 2010MNRAS.404.1111G, 2010ApJ...710..903M, 2013MNRAS.428.3121M, 2018MNRAS.477.1822M, 2010ApJ...717..379B, 2013ApJ...777L..10B, 2019MNRAS.488.3143B, 2013ApJ...771...30R, 2017ApJ...840...34S, 2017MNRAS.470..651R, 2012ApJ...752...41Y}. These models span the entire halo mass range ($M_{\rm halo} \sim 10^9$--$10^{15} M_\odot$) and can explore redshift evolution. By construction, however, they are indirect, rely on assumptions regarding scatter and star formation efficiency, and may be biased toward certain galaxy populations, such as red or quenched systems.

X-ray observations of the hot intracluster and intragroup medium provide an alternative route to direct halo mass estimation. Indeed, the temperature of the hot gas is considered a very accurate proxy of the halo mass {in virialized systems} due to the predicted and observed tight correlation between temperature and mass \citep{2004ApJ...617..879L,2013ApJ...778...14G,2018AstL...44....8K,2019A&A...631A.175E,2016MNRAS.455..258C,2025A&A...704A.110C,2022PASJ...74..175A, 2018MNRAS.474.3009D}. This approach is robust for massive halos ($M_{\rm halo} \gtrsim 10^{14} M_\odot$), where the hot gas dominates, but becomes increasingly limited in lower-mass systems, where X-ray emission is generally too faint for reliable measurements of individual halos for the most of the sample. Indeed, detecting low-mass halos ($10^{12} < M_{\rm halo} < 10^{14} M_{\odot}$) in X-rays remains challenging. As a result, obtaining reliable halo mass estimates from X-ray observations is difficult for most low-mass groups on an individual basis.

Stacking offers a powerful method of overcoming the limitations in measuring individual halo properties. By combining X-ray spectra from galaxy groups selected through different techniques, it becomes possible to recover the average gas temperatures and total X-ray luminosity \citep{2025A&A...700A.167T, Zheng_2023}. Wide-field surveys such as the eROSITA All-Sky Survey (eRASS1; \citealt{2024A&A...682A..34M}) and the upcoming eRASS:4 do not provide complete samples of galaxy groups across large volumes, due to strong selection effects \citep{paola_24, paola_stacking_magneticum, ilaria_lightcone}. However, these surveys enable the stacking of large numbers of faint systems, significantly boosting the signal-to-noise ratio and enabling the study of group-scale halos that are too faint to detect individually.

In this paper, we use spectral stacking to measure gas temperatures and infer halo masses across more than three decades in halo mass. This approach allows us, for the first time, to observationally trace the SHMR down to $M_{\rm halo} \sim 10^{12}\,M_{\odot}$, with stellar and halo masses independently derived from optical and X-ray data, respectively. In contrast to previous studies, which focused on limited mass ranges or required individually detected systems, we present average measurements for a large, optically selected population of galaxy groups spanning a broad range in halo mass. The efficiency, completeness, and purity of the optical group catalog used {as a parent sample} \citep{yang_galaxy_2007} have been extensively validated in previous works \citep{paola_stacking_magneticum, ilaria_lightcone, ilaria_opticallightcone, 2025A&A...695C...1M}. We stack this group sample using data from eRASS1 to provide the first observational constraints on the SHMR from Milky Way--mass galaxy groups up to massive clusters.

Throughout this paper, we adopt a flat $\Lambda$ cold dark matter cosmology with ${\Omega_{\mathrm{m}}} = 0.27$ and $H_0 = 70\;\mathrm{km}\;\mathrm{s}^{-1}\;\mathrm{Mpc}^{-1}$, {and use a Chabrier initial mass function (IMF; \citealt{2003PASP..115..763C})}. {$R_{\Delta}$ is the radius within which the mean density of the halo is $\Delta$ times higher than the critical density of the Universe, and $M_{\Delta}$ is the total mass enclosed within this radius.} {We refer to the halo masses based on the optical proxy as $M_{\Delta, \mathrm{opt}}$ and to the X-ray derived halo masses as $M_{\Delta, \mathrm{X}}$.}
 
\section{Analysis framework}

\subsection{{Identifying halos and their central galaxy}}\label{sec:sample}

Galaxy groups were selected from the catalog of \cite{yang_galaxy_2007}, {covering the low-redshift range up to $z<0.2$}. The reliability of this optical selection algorithm in identifying galaxy groups and their central galaxies was assessed by \cite{ilaria_opticallightcone}, who compared its performance against a simulated galaxy light cone based on the {\sc Magneticum}\footnote{\url{http://www.magneticum.org/index.html}} \citep{2016MNRAS.463.1797D, 2025arXiv250401061D} simulation. The algorithm presented in \cite{yang_galaxy_2007} achieves {90\% completeness {in the reconstruction of the simulated sample} at $\sim10^{13.5}\,M_{\odot}$, remaining above 60\% down to $\sim10^{12}\,M_{\odot}$,} with high purity ($>93\%$). Moreover, the identification of central galaxies is highly efficient, with a success rate of 97\%. {We use the total stellar luminosity from the \cite{yang_galaxy_2007} catalog as a ${M_{200,\mathrm{opt}}}$ halo mass proxy via a mass-to-light ratio. Its intrinsic scatter of $\sigma_{{M_{200,\mathrm{opt}}}}\sim0.25\,\mathrm{dex}$ on average \citep{ilaria_opticallightcone} is smaller than our adopted bin widths at fixed halo mass, and hence ensures the minimal contamination from systems of lower or higher mass in our binning analysis, as quantified in \cite{paola_stacking_magneticum}.} {Previous analysis of the skewness and kurtosis of a similar sample indicates that, on average, the systems tend to be virialized \citep{paola_profiles}.}

We followed the methodology of \cite{2025A&A...700A.167T} for group selection, X-ray data extraction, and spectral stacking. {That work used the \cite{yang_galaxy_2007} catalog with a halo mass cut of ${M_{200,\mathrm{opt}}} > 10^{13}\,M_{\odot}$ and additional selection criteria to remove point source contamination and fragmentation.} 

{Here we expanded the \cite{2025A&A...700A.167T} sample through the following steps:
\begin{enumerate}
    \item We started from the full \cite{yang_galaxy_2007} catalog of 472,504 systems with redshift $z<0.2$;
    \item We restricted the sample to the eRASS1 footprint, resulting in 205,903 systems;
    \item For each system, $R_{500}$ was estimated from $M_{180}$ (assuming $M_{180} \sim M_{200}$) using the {\sc Colossus} package \citep{2018ApJS..239...35D}, through the mass-concentration relation of \cite{2019ApJ...871..168D} and a Navarro-Frenk-White density profile \citep{1997ApJ...490..493N}. We then removed systems whose $R_{500}$ overlaps with other groups or clusters from the same catalog, leaving 163,735 isolated systems;
    \item We further removed systems whose $R_{500}$ aperture contains point sources detected in eRASS1 (defined as $\mathrm{EXT\_LIKE} = 0$ and $\mathrm{DET\_LIKE} \geqslant 6$), leaving 156,990 systems;
    \item Finally, we selected systems with either a BCG stellar mass of $M_{\star,\mathrm{BCG}} > 10^{10.3}\,M_{\odot}$, using stellar masses from the GSWLC-X2 catalog \citep{Salim} for the BCGs defined in \cite{yang_galaxy_2007}, or a halo mass of $M_{200,\mathrm{opt}} > 10^{11.7}\,M_{\odot}$.
\end{enumerate}
This selection procedure produced a final sample of 112,699 systems in the range of $10^{11.7}\,M_{\odot} < M_{200,\mathrm{opt}} < 10^{14.1}\,M_{\odot}$ and $10^{10.3}\,M_{\odot} < M_{\star,\mathrm{BCG}} < 10^{11.6}\,M_{\odot}$.}

\begin{figure}
\centering
\includegraphics[width=1\hsize]{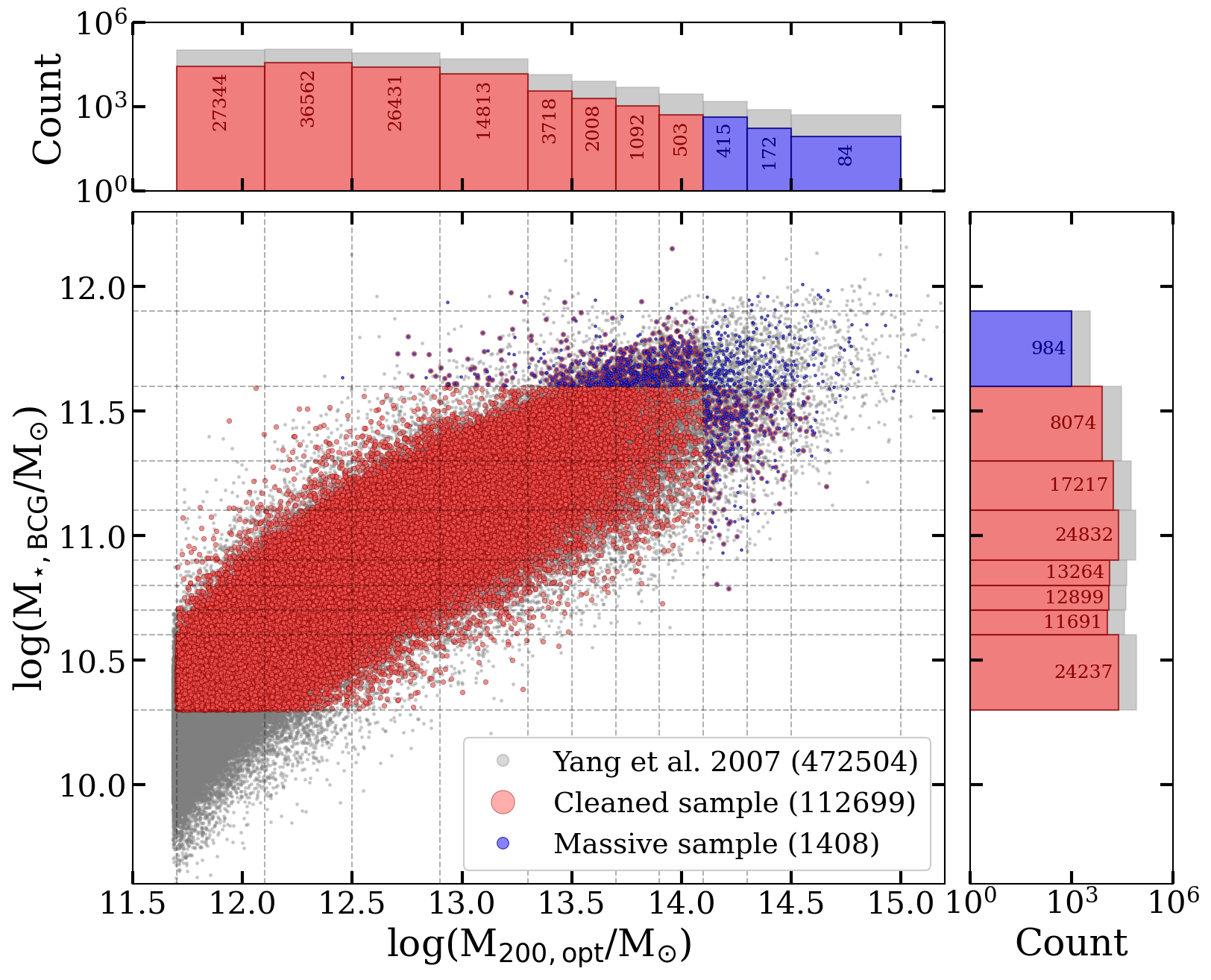}
  \caption{{Distribution of the parent \cite{yang_galaxy_2007} catalog (gray), the final cleaned sample (red), and the massive subsample (blue) in the $M_{200,\mathrm{opt}}$--$M_{\star,\mathrm{BCG}}$ plane. Dashed lines mark the bin edges used throughout this work. The top and side panels show the corresponding one-dimensional distributions in $M_{200,\mathrm{opt}}$ and $M_{\star,\mathrm{BCG}}$, with the number of sources in each bin indicated for the subsamples used in the further analysis.}
\label{fig:sample_selection}}
\end{figure}

\begin{figure}
\centering
\includegraphics[width=1\linewidth]{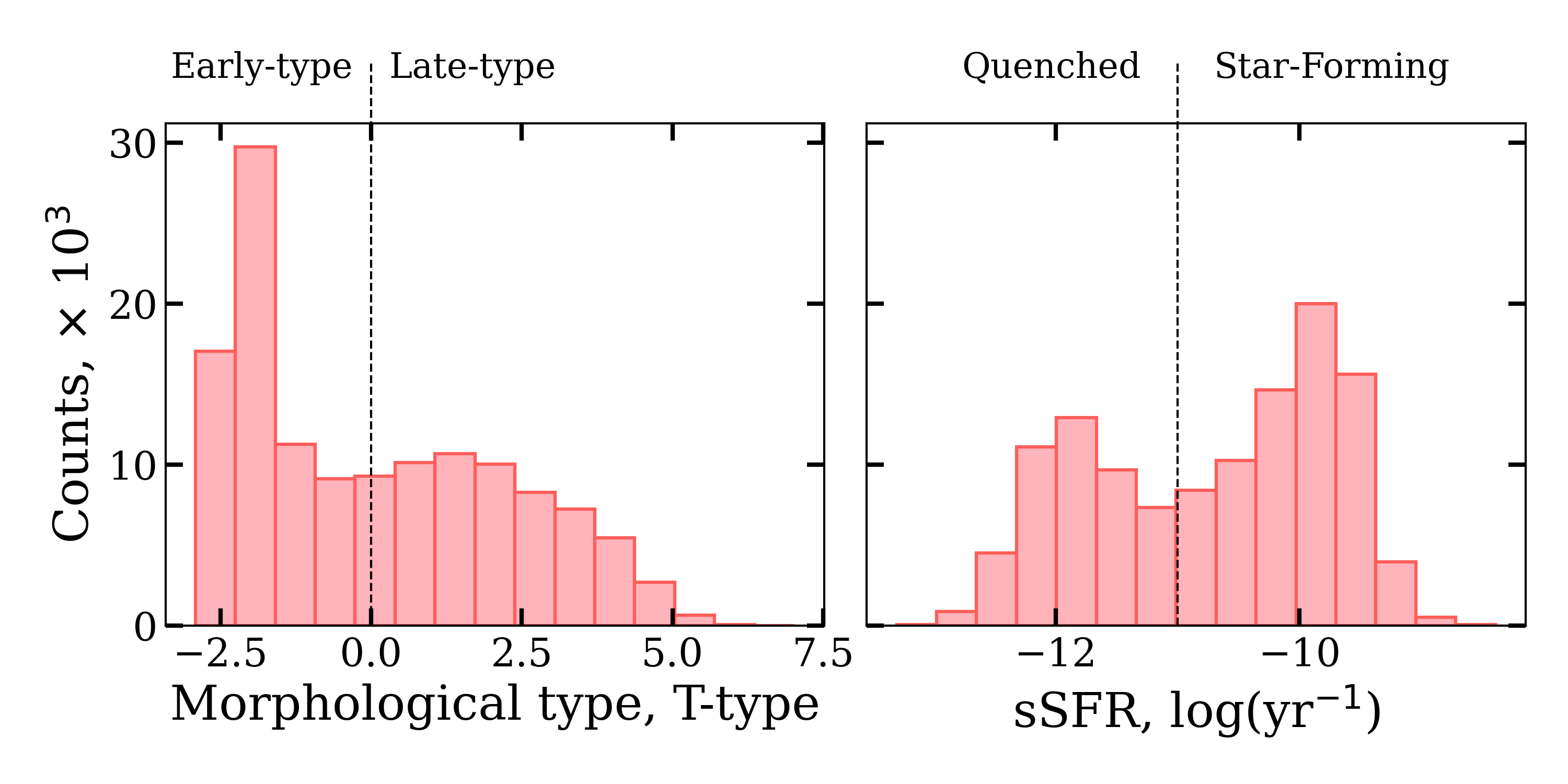}
\caption{{Distribution of morphological type (T-type; \textit{left}) and sSFR (\textit{right}) for the BCGs in our sample. The dashed lines mark the boundaries separating early-type galaxies from late-type ones (T-type $= 0$) and quenched galaxies from star-forming ones (sSFR $= 10^{-11}\,\mathrm{yr}^{-1}$).}}
\label{fig:bcg_properties}
\end{figure}

{While the high performance of the \cite{yang_galaxy_2007} group finder is confirmed for systems with ${M_{200,\mathrm{opt}}} < 10^{14}\,M_{\odot}$, as discussed above, its purity and completeness at the high-mass end, on cluster scales, may be affected by fragmentation \citep{ilaria_lightcone}. To mitigate this effect, we constructed the high-mass end of the sample (${M_{200,\mathrm{opt}}} > 10^{14.1}\,M_{\odot}$ and ${M_{\star,\mathrm{BCG}} < 10^{11.6}\,M_{\odot}}$) differently from the rest of the sample.} {Starting from the 2,162 systems in this mass range, we removed those whose $R_{500}$ aperture contains point sources detected in eRASS1, leaving 1,416 systems; we did not apply the isolation criterion in this mass range, since massive clusters can be fragmented into a main halo and several lower-mass neighboring halos, which would naturally fall within the $R_{500}$ of the main system. We then cross-matched these 1,416 systems within $R_{500}$ with the eRASS1 X-ray-selected cluster catalog \citep{2024A&A...685A.106B} with a $4'$ radius, and removed 8 systems with a match and an eRASS1 redshift of $\mathrm{BEST\_Z} > 0.2$, since these correspond to background X-ray sources unrelated to our $z < 0.2$ sample. This procedure resulted in a final sample of 1,408 clusters.}

{Figure~\ref{fig:sample_selection} shows the distribution of the parent sample, the final cleaned sample, and the massive subsample, together with the binning scheme and the source counts in each bin used throughout this work. We also assessed the properties of the BCGs in our sample, using specific star formation rate (sSFR) values from the GSWLC catalog and morphological T-type classifications from \cite{2018MNRAS.476.3661D}. As shown in Fig.~\ref{fig:bcg_properties}, our sample spans a broad range of BCG morphologies and evolutionary stages, including both early-type and late-type galaxies, as well as both quenched and star-forming systems.}

For each group, source and background X-ray spectra were extracted from the eRASS1 public event lists. Source spectra were measured within $R_{500}$, centered on the BCG coordinates, while background spectra were extracted from annular regions spanning $1.5$--$1.8\,R_{500}$, {with the annulus area matched to the area of the source region to} ensure consistent background subtraction.

{To test the robustness of the results against intrinsic scatter and measurement uncertainties,} we performed the stacking analysis using two independent binning schemes: one based on BCG stellar mass and the other based on the optical halo mass proxy. {This bidirectional binning approach allowed us to assess whether any potential bias introduced by binning along a single axis, expected when using a mass proxy with intrinsic scatter \citep{2014MNRAS.441.3562E}, is subdominant compared to the statistical uncertainties in the derived $M_{\star,\mathrm{BCG}}$--$T_{\mathrm{X}}$ and $M_{\star,\mathrm{BCG}}$--$M_{\mathrm{halo}}$ relations.} By comparing the resulting relations, we could verify the consistency of the observed trends, independent of the choice of binning variable.

The spectra in each bin were stacked to produce an average spectrum representative of that mass range. {After stacking, the background spectra were subtracted from the corresponding source-plus-background spectra before spectral modeling. As demonstrated in \cite{2025A&A...700A.167T}, this background-subtraction approach gives results consistent with those obtained from the simultaneous modeling of the source-plus-background and background spectra.} Bin definitions and the corresponding stacked {background-subtracted} spectra are provided in Appendix~\ref{ap:spectra}. Additional details on the selection process, data extraction, stacking procedure, and validation tests can be found in \cite{2025A&A...700A.167T}.

\subsection{Deriving gas temperatures and halo masses\label{sec:fitting}}

We fit the {background-subtracted} stacked spectra in each bin using the Sherpa (v4.17.0) Python package \citep{sherpa}, modeling the emission with multiple components: an intracluster medium (ICM) component, a power-law component representing unresolved AGNs, and Galactic line-of-sight absorption. {Since the {X-ray binary stars} contribution is low \citep{2025A&A...700A.167T} and has a similar power-law shape that may lie within the error budget of the AGN component, we did not model it separately.} 

The ICM emission was described using the {\sc gadem}\footnote{\url{https://heasarc.gsfc.nasa.gov/xanadu/xspec/manual/XSmodelGadem.html}} model, which represents a multi-temperature plasma with a Gaussian distribution of emission measure, implemented as a combination of multiple {\sc mekal}\footnote{\url{https://heasarc.gsfc.nasa.gov/docs/xanadu/xspec/manual/XSmodelMekal.html}} components \citep{1985A&AS...62..197M,1986A&AS...65..511M,1995ApJ...438L.115L}. {For each bin,} the initial predictions of the mean temperature and Gaussian width {were derived as follows: the optical halo masses, $M_{200,\mathrm{opt}}$, from the \cite{yang_galaxy_2007} group catalog of all sources in the bin were converted to temperatures using the $M$--$T_\mathrm{X}$ relation of \cite{2025A&A...700A.167T}, and the mean and standard deviation of the resulting distribution were used as the initial values, respectively.} {The Gaussian width accounts for the physical diversity of sources in the stack: since the stacked spectrum for each bin represents an ensemble of sources with intrinsically different temperatures due to the halo mass distribution within the bin, a nonzero width is required to reproduce the shape of the ensemble spectrum.} During the fitting process, the mean temperature was allowed to vary, while the width of the temperature distribution was fixed {to reduce the number of free parameters and to limit degeneracies in the fit. Allowing the width to vary increases the statistical uncertainties and introduces degeneracies or spurious local minima, without significantly changing the best-fit width. For this reason, we fixed the width to ensure a more robust fit (see Fig.~\ref{apf:tsigma} for a comparison).} {Following the recommendations of \cite{abundances_1}, \cite{gastadello_review}, \cite{abundances_xcop}, and \cite{abundances_suzaku}, the metal abundance in each bin was fixed to $0.3\,Z_\odot$ }{ using the abundance table from \cite{1989GeCoA..53..197A}}. {For more details on the impact of different metal abundance assumptions, see Sect. \ref{sec:abundances}.}

The power-law component accounts for unresolved AGNs, which are particularly important in the low-mass bins \citep[see also][]{paola_stacking_magneticum,2025A&A...700A.167T}. To limit the number of free parameters, the photon index was fixed to the average AGN value $\Gamma = 1.95$ \citep{1994MNRAS.268..405N}. {Because the stacked spectra, especially at low halo masses, have limited signal-to-noise, leaving both the ICM and AGN normalizations free can lead to degeneracies between the fit parameters. We therefore constrained the relative normalization of the two components using the ICM-to-AGN flux ratio predicted by the {\sc Magneticum} hydrodynamical simulation \citep{2025A&A...700A.167T}. This ratio is a function of halo mass. Since $M_{500,\mathrm{X}}$ was not known a priori and was instead inferred from the fit temperature through the \mbox{$M$--{$T_\mathrm{X}$}} relation, the normalization constraint was implemented as a temperature-dependent relation. During the fitting process, for each temperature iteration, the corresponding halo mass was computed, and the ICM-to-AGN normalization ratio was updated accordingly.} The ICM component dominates in the {regions of the Fe-L and Fe-K complexes}, while the power law contributes primarily at the edges of the eROSITA effective area (below $\sim 0.5$~keV and above $\sim 1.5$~keV). Line-of-sight absorption was included, with initial $n_{\rm H}$ values taken from the {\sc ftools} database and {allowed to vary with} lower and upper bounds of $0.1$--$0.7 \cdot 10^{22}\,{\rm cm}^{-2}$. {Therefore, the spectral model has three free parameters: the mean temperature, $T_\mathrm{X}$, the ICM component normalization, and the hydrogen column density, $n_\mathrm{H}$.}

From the fit spectra, we derived the {mean} temperatures for each BCG-stellar- and halo- mass bin (see Appendix~\ref{ap:spectra}) and converted them to X-ray-based halo masses, $M_{500,\mathrm{X}}$, using the \mbox{$M$--{$T_\mathrm{X}$}} relation \citep{lovisari_relation}. Uncertainties {on the temperature measurements and, consequently, on $M_{500,\mathrm{X}}$} were estimated via the bootstrap procedure described and validated in \cite{2025A&A...700A.167T}.  {$M_{500,\mathrm{X}}$ was then converted to ${M_{200,\mathrm{X}}}$ using the {\sc Python} package {\sc Colossus} \citep{2018ApJS..239...35D}. {Halo concentrations were computed individually for each halo as a function of halo mass and redshift using the mass-concentration relation \citep{2019ApJ...871..168D}.} A comparison between the X-ray- and optically based halo mass estimates shows a high consistency between the two, supporting the reliability of the mass measurements (see Appendix~\ref{apf:mh_mh}).} {A comparison between the average temperature of individual spectra in the highest-mass bin and the corresponding stacked result is provided in Appendix~\ref{ap:individual}.}

\section{Linking {BCG} stellar mass and halo temperature}

\begin{figure}
\centering
\includegraphics[width=1\hsize]{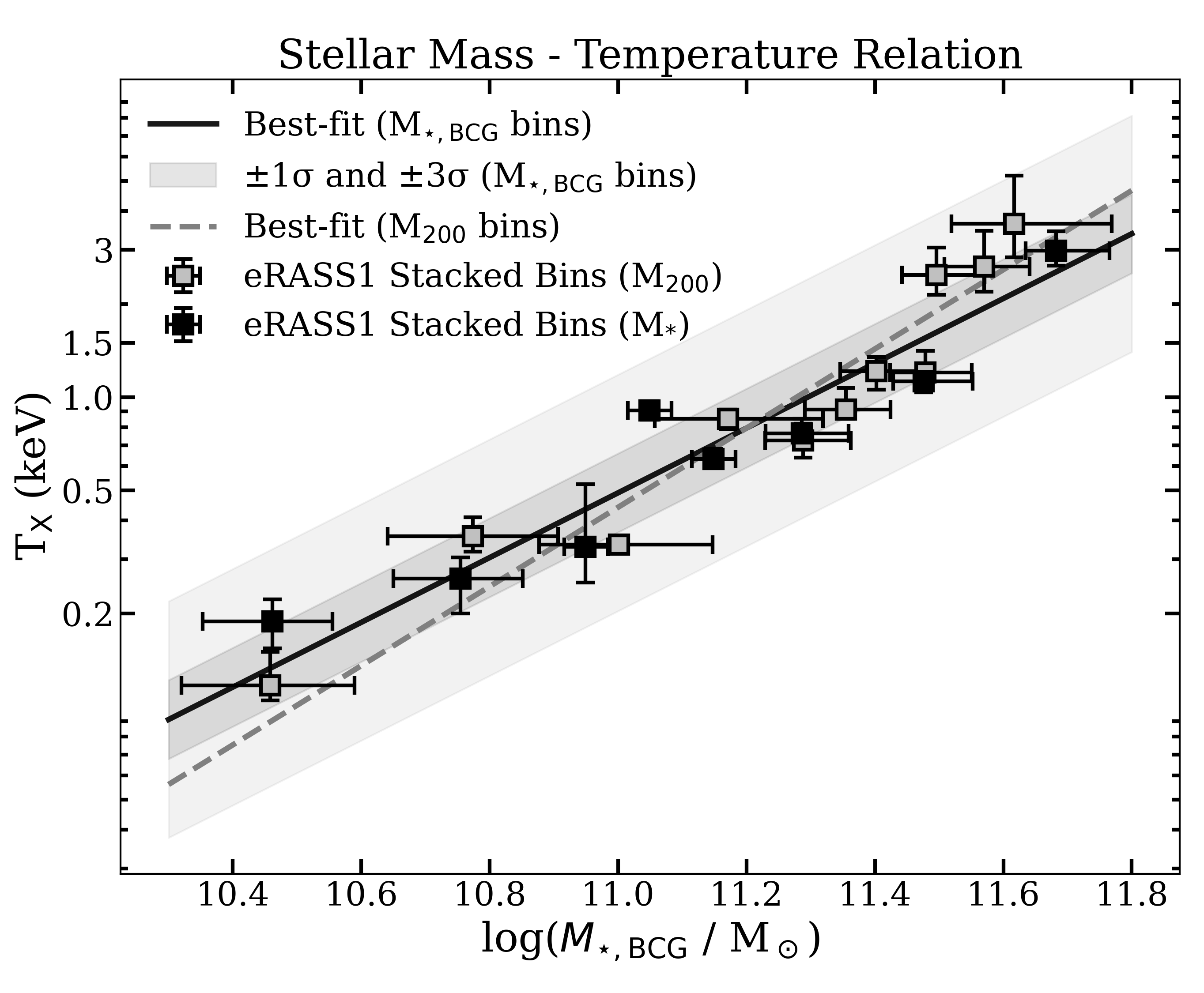}
  \caption{Stacked X-ray temperatures within $R_{500}$ as a function of BCG stellar mass. Black points correspond to bins defined by stellar mass, while gray points correspond to bins defined by {optically based halo mass \citep{yang_galaxy_2007}; for both binning schemes,} the average BCG stellar mass of each sub-bin is plotted on the $x$ axis. Error bars on the $y$ axis indicate $1\sigma$ uncertainties derived from bootstrap resampling, while error bars on the $x$ axis represent the {1$\sigma$ interval around the mean. Best-fit relations for both binning schemes are shown as solid black and dashed gray lines. The shaded area corresponds to the 1$\sigma$ and 3$\sigma$ uncertainties of the best fit.} \label{fig:mstar_t}}
\end{figure}

Figure~\ref{fig:mstar_t} shows the stacked X-ray temperatures as a function of BCG stellar mass. The relation is presented for both stellar-mass and halo-mass binning; {in the latter case, the binning is based on optically derived halo masses from \cite{yang_galaxy_2007}, and} we plot the average stellar mass of the BCG within each sub-bin.

We recover well-defined temperatures across all bins, revealing a smooth, approximately log-linear increase in temperature with stellar mass, from $M_{\star,\mathrm{BCG}} \approx 10^{10.3}\,M_\odot$ to $M_{\star,\mathrm{BCG}} \approx 10^{11.8}\,M_\odot$. The results obtained from stellar-mass and halo-mass binning agree within the uncertainties, indicating that the choice of binning does not strongly affect the overall trend. However, as shown in Appendix~\ref{ap:spectra}, spectra corresponding to the halo-mass bins at the low-mass end exhibit fewer degrees of freedom and systematically higher $\chi^2$ values in the spectral modeling. This likely reflects inaccuracy in the halo mass proxy due to incompleteness in either the group membership or the calibration of the mass proxy.

The observed correlation confirms the expected connection between the stellar content of BCGs and the depth of the halo potential, providing a direct link between galaxy properties and halo properties. {This result indicates the stellar mass of BCGs as a useful proxy for halo mass across the group-to-cluster mass regime, although the intrinsic scatter may limit its precision as a mass estimator.}

{We fit the \mbox{$M_{\star,\mathrm{BCG}}$--{$T_\mathrm{X}$}} relation using orthogonal distance regression (ODR) implemented via the {\sc scipy.odr} module from the {\sc scipy} {\sc Python} package, accounting for uncertainties in the temperature axis. The best fit of the relation was obtained separately for the two binning schemes: }

\begin{equation}
\log_{10} \left( \frac{T_{\mathrm{X}}}{1\;\mathrm{keV}} \right) = {1.28\pm0.13}\cdot \log_{10} \left(\frac{M_{\star,\mathrm{BCG}}}{M_{\odot}}\right) - {14.43\pm1.47},
\end{equation}
for the ${M_{200,\mathrm{opt}}}$ binning scheme, and
\begin{equation}
\log_{10} \left( \frac{T_{\mathrm{X}}}{1\;\mathrm{keV}} \right) = {1.05\pm0.15}\cdot \log_{10}\left(\frac{M_{\star,\mathrm{BCG}}}{M_{\odot}}\right) - {11.82\pm1.61},
\label{eq:mstar_t}\end{equation}
for the $M_{\star,\mathrm{BCG}}$ binning scheme.

{Since the difference between the two fit relations is minor (within 1.4$\sigma$), {the binning-induced bias is negligible compared to the current statistical uncertainties, and both subsamples follow a consistent $M_{\star,\mathrm{BCG}}$--$T_{\mathrm{X}}$ trend.}}

\section{Stellar-to-halo mass relation}

\subsection{First observational view of the relation}

\begin{figure}
\centering
\includegraphics[width=1\hsize]{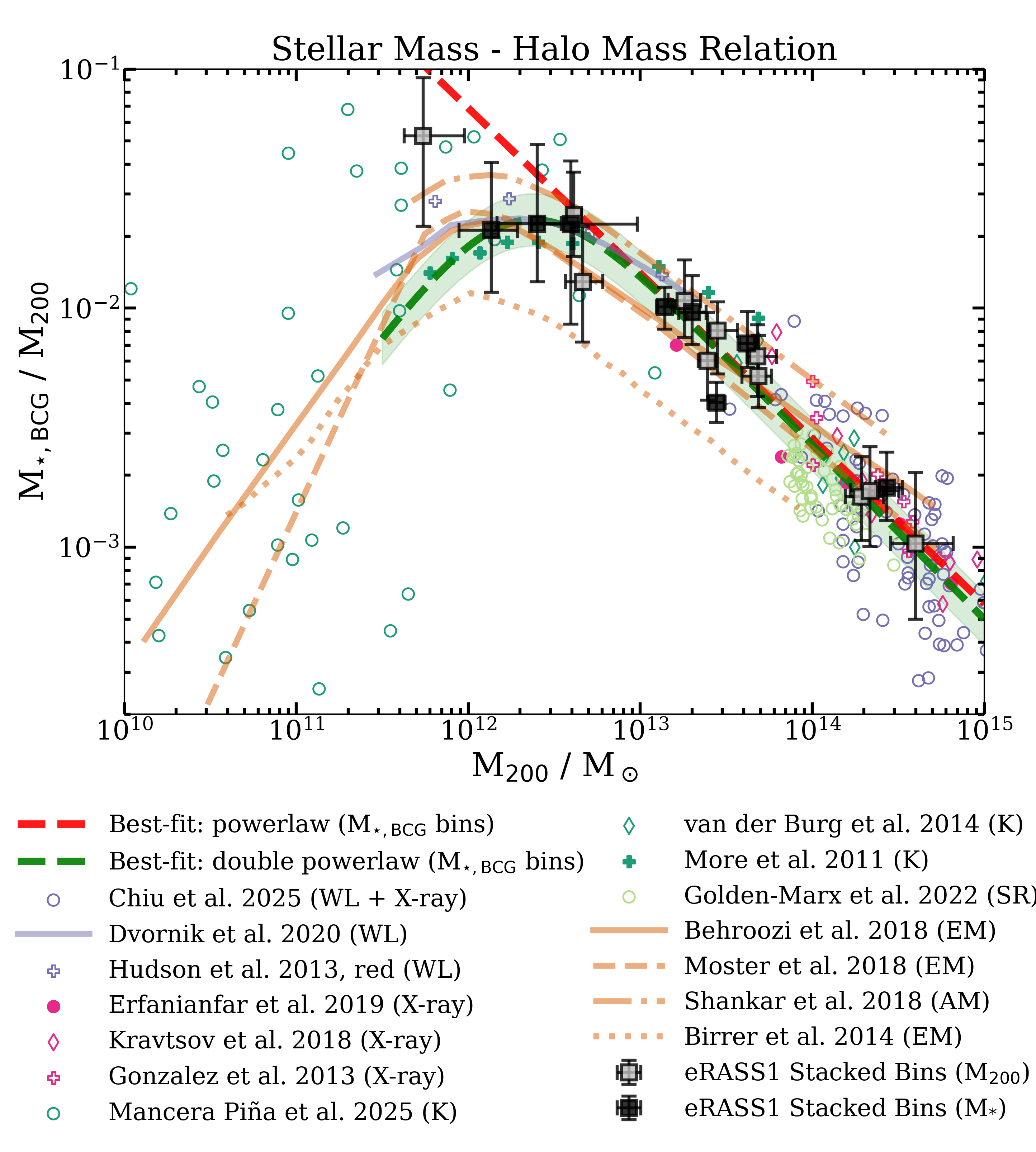}
  \caption{BCG stellar-to-halo mass ratio as a function of halo mass. The meaning of the symbols is as in Fig.~\ref{fig:mstar_t}. {Best-fit relations are shown as dashed green and red lines; the shaded region indicates the $1\sigma$ uncertainty of the double power-law best-fit relation.} Literature measurements are color-coded by halo mass estimation method as indicated in the legend: {violet for WL, red for X-ray, green for kinematical and dynamical studies (K), and light green for scaling relations (SR); for the literature data, filled points indicate stacking results, while empty points show results for individual sources. Orange lines correspond to semi-analytical predictions from abundance matching (AM) and empirical modeling (EM).}
  \label{fig:shmr_obs}}
\end{figure}

Figure \ref{fig:shmr_obs} presents the SHMR derived from the X-ray-based halo masses. Our measurements cover the range ${M_{200,\mathrm{X}}} \approx 10^{12}$--$10^{15} M_\odot$, providing the first observational SHMR spanning this full mass range with halo masses directly inferred from X-ray gas temperatures. Error bars indicate $1\sigma$ uncertainties derived via bootstrap resampling of the stacked spectra.

Around the expected characteristic turnover, we observe a plateau in the stellar-to-halo mass ratio, $M_{\star,\mathrm{BCG}}/{M_{200,\mathrm{X}}}$, which remains approximately constant within the uncertainties. This feature indicates that the turnover is well captured in our data, corresponding to the peak in star formation efficiency near $M_{200} \sim 10^{12} M_\odot$ \citep{Yang_2009, 2009ApJ...696..620C, 2010ApJ...717..379B, 2010MNRAS.404.1111G}. Above this halo mass, the ratio declines, reflecting the increasing importance of AGN feedback in suppressing in situ star formation \citep{ 1998A&A...331L...1S, 2007ARA&A..45..117M, 2012ARA&A..50..353K}. 

Extending the SHMR to lower halo masses to fully map the turnover and subsequent decline is {currently} unfeasible, even with deeper wide-field surveys such as eRASS:4. At ${M_{200,\mathrm{X}}} \approx 10^{11} M_\odot$, the expected X-ray temperature is $T_X = 0.036 \pm 0.009~\mathrm{keV}$, corresponding to halos whose gas content is negligible. In this regime, the majority of the emission lies outside the sensitive energy range of current X-ray observatories (eROSITA, {\textit{XMM-Newton}}, and {\textit{Chandra}}), making direct detection and reliable spectral modeling impossible for such halos. {For the same reason, we note that the temperatures in the lowest-mass bins should be interpreted with caution, as they lie close to the low-energy sensitivity limit, which may introduce additional systematic uncertainties.}

{As for the \mbox{$M_{\star,\mathrm{BCG}}$--{$T_\mathrm{X}$}} relation, 
{the stacked points from the two binning schemes show good agreement. To fit the SHMR, we selected only the $M_{\star,\mathrm{BCG}}$ binning, which provides smaller uncertainties on the derived temperatures ({see} Appendix \ref{tab:spectral_fit_results})}. The best fit of the relation was obtained with ODR in two ways. First, we simply fit a power law to the data at halo masses larger than the knee $M_{\mathrm{peak}}\approx 4\cdot10^{12} M_{\odot}$. The best fit is}
\begin{equation}
{
\log_{10} \left( \frac{M_{\star,\mathrm{BCG}}}{{M_{200,\mathrm{X}}}} \right) = {-0.69\pm0.05}\cdot \log_{10}\left(\frac{{M_{200,\mathrm{X}}}}{M_{\odot}}\right) - {7.13\pm0.77},}
\end{equation}

{Second, we fit the entire relation with a double power law following the prescription from \cite{2019MNRAS.488.3143B}, but without a Gaussian component:}
\begin{equation}
\log_{10} \left( \frac{M_{\star,\mathrm{BCG}}}{M_0} \right) = \epsilon - \log_{10} \left( 10^{-\alpha x} + 10^{-\beta x + \kappa} \right), \label{eq:behroozi}
\end{equation}
\begin{equation}
x =\log_{10} \left( \frac{{M_{200,\mathrm{X}}}}{M_{0}} \right)
.\end{equation}

{\noindent Here we fixed the faint end slope $\alpha=1.963$ and scale $\log_{10}(M_0/M_{\odot})=12.035$ following the best-fit value of \cite{2019MNRAS.488.3143B} and leaving free parameters constraining the bright end slope, $\beta$, and normalization, $\epsilon$.}
The best fit values are\begin{equation*}
{
\begin{aligned}
\beta &= 0.27 \pm 0.10, \\
\epsilon &= -1.59 \pm 0.09, \\
\kappa &= -0.45 \pm 0.19.
\end{aligned}
}
\end{equation*}
{
Both best-fit relations are shown in Fig.~\ref{fig:shmr_obs}. While a single power law describes the data well over the fit mass range, the double power law not only reproduces the data {to the right of the knee}, but also accurately captures the shape of the knee for the $M_{\star}$ binning scheme.}

\subsection{Comparison with previous results}\label{sec:literature_comp}

Figure~\ref{fig:shmr_obs} compares our SHMR measurements with a selection of literature results (not all studies are shown to preserve figure clarity; additional references are discussed in the text). {For the comparison, we converted all the literature stellar masses to the Chabrier IMF \citep{2003PASP..115..763C} if a different IMF was used.} Overall, the results are consistent within uncertainties across various methods, despite significant scatter. 

In earlier sections we discussed how differences in halo mass estimation can impact the SHMR {(see Sect. \ref{sec:intro})}. However, variations in BCG stellar mass estimates also play a role. {While all the literature data were converted to the same IMF, other assumptions, such as star formation history, dust, and metallicity, can also affect the stellar mass estimates and contribute to the scatter between studies.}

{Moreover, even among spectral energy distribution-based approaches with consistent model assumptions, the choice of spatial aperture for measuring stellar mass can significantly influence the resulting SHMR.} {We adopt the stellar masses from \cite{Salim}, which were derived using a broad range of photometric bands (from near-UV to far-infrared) and a PDF-matching method.} In contrast, studies that rely on significantly larger apertures, which can include contributions from intracluster light (ICL), tend to produce flatter SHMR slopes at the high-mass end due to systematically higher BCG stellar masses at the cluster scale \citep{2018AstL...44....8K, 2018MNRAS.474.3009D}. {The ICL contribution may account for up to $\sim25\%$ of the total stellar mass of cluster members \citep{2025MNRAS.537.3954M, 2026A&A...708A.124M}.} In this work, we do not explore the effect of larger apertures for galaxy groups. {While ICL can measurably affect BCG stellar masses in clusters, its contribution at the group scale is expected to be smaller and remains poorly constrained, and we therefore do not include it in our analysis.} To maintain consistency and avoid aperture-related biases, we base our SHMR on stellar masses that exclude intracluster and intragroup light, and compare only with measurements derived using similar apertures. Specifically, for the \cite{2018AstL...44....8K} data shown in Fig.~\ref{fig:shmr_obs}, we adopt the values measured within 50~kpc, following the approach of \cite{2025A&A...704A.110C}.

At the high-mass end, our results show excellent consistency with individual eRASS1 cluster detections from \cite{2025A&A...704A.110C}, which naturally align with the best-fit SHMR. Similar agreement is found with other studies based on X-ray--derived halo masses \citep{2019A&A...631A.175E, 2013ApJ...778...14G, 2018AstL...44....8K} and with halo mass estimates inferred from optical scaling relations \citep{2014A&A...561A..79V, 2018ApJ...860....2G}.

At lower halo masses, we find good agreement with observational constraints on the location of the SHMR knee. {In the vicinity of the knee and toward higher halo masses, the SHMR normalization reported by \cite{2020A&A...642A..83D}, based on weak-lensing-calibrated halo masses in the GAMA survey area, is consistent with our results. Similar consistency is found in studies relying on optical mass proxies over more limited halo mass ranges, such as \cite{2011MNRAS.410..210M}. At the lowest halo masses probed in this study, {closer to} the regime of individual galaxy halos, our measurements connect smoothly to the high-mass end of the galactic-halo sample \citep{2025A&A...699A.311M, 2021ApJ...923...35M, 2024ApJ...971...69D}.}

Among semiempirical models, most \citep{2010ApJ...710..903M, 2013ApJ...777L..10B, 2019MNRAS.488.3143B, 2013MNRAS.431..648W, 2015MNRAS.450.1604L, 2017MNRAS.470..651R,2017ApJ...840...34S} are consistent with our measurements within $1\sigma$. The only exception is the work of {\cite{2014ApJ...793...12B}, which shows a moderate offset toward lower stellar-to-halo mass ratios.}

\subsection{Impact of gas metallicity}\label{sec:abundances}

\begin{figure}
\centering
\includegraphics[width=1\hsize]{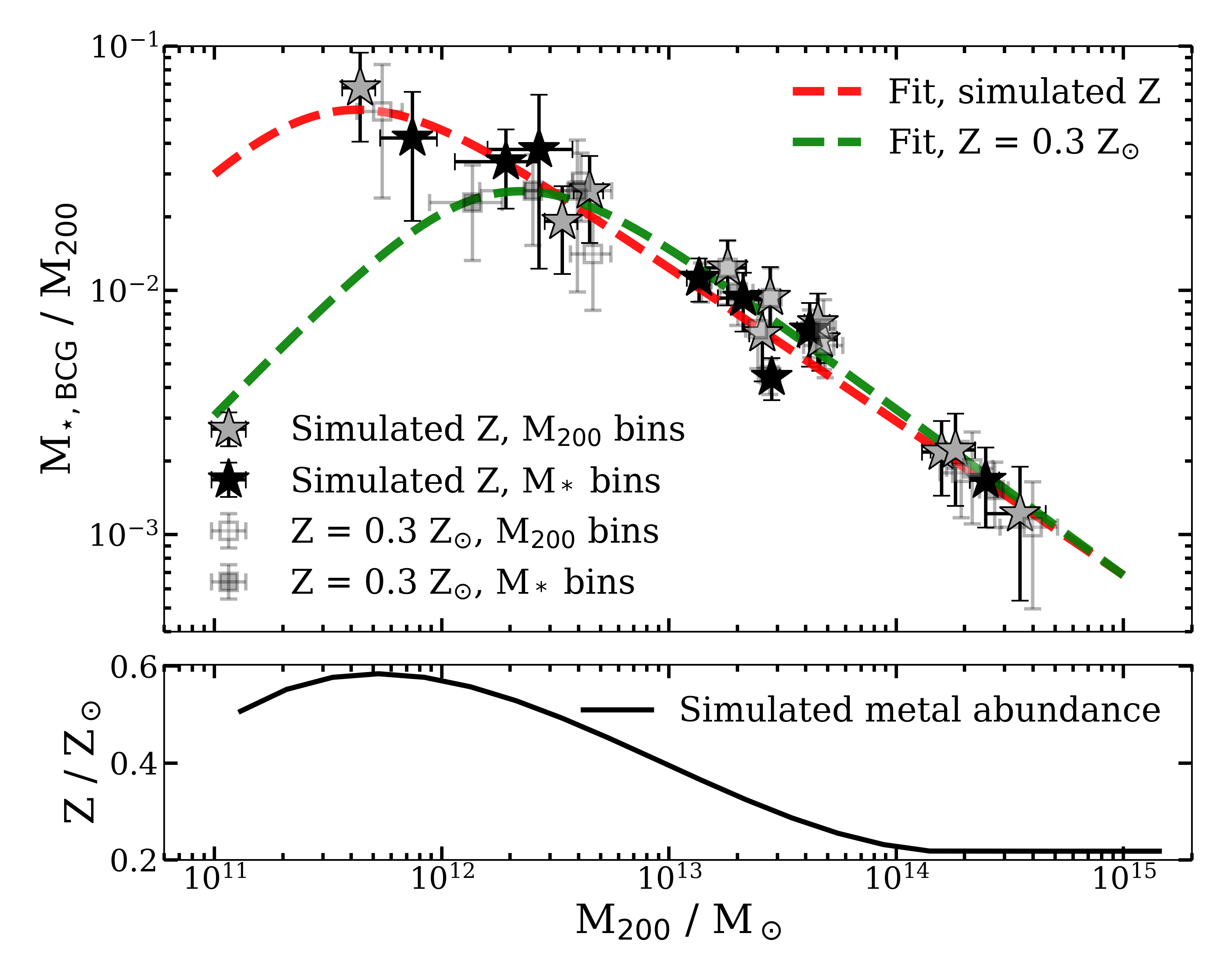}
\caption{{\textit{Top:} Stellar-to-halo mass ratio as a function of halo mass for different assumptions on the gas metal abundance. Semitransparent squares show the results of the main analysis, identical to those presented in Fig.~\ref{fig:shmr_obs}. Star symbols correspond to the same stacked spectra modeled using simulated metal abundances from the bottom panel. Dashed lines indicate the best-fitting double power-law relations for each case, following Eq.~\ref{eq:behroozi}. \textit{Bottom:} Mean gas metal abundance as a function of halo mass derived from the {\sc Magneticum} simulation and adopted for the spectral modeling shown in the top panel.}}
\label{fig:shmr_0p3}
\end{figure}

{We explored the impact of the assumed gas-phase metal abundance in the spectral modeling on the derived SHMR. In our main analysis, the metal abundance in each halo mass bin was fixed to a constant value of $0.3\,Z_\odot$ (see Sect. \ref{sec:fitting}). To assess the sensitivity of our results to this assumption, we performed an alternative analysis in which the metal abundance was allowed to vary with halo mass, adopting the mean metal abundance within $R_{500}$ derived from the {\sc Magneticum} hydrodynamical simulation for each bin{, since observationally the gas metallicity for the lowest mass bins is poorly constrained}. {The metal abundances were computed for the hot gas only, weighted by particle mass.} The bottom panel of Figure~\ref{fig:shmr_0p3} shows the resulting mean metal abundance as a function of halo mass, which was used for the spectral modeling in this comparison analysis.}

{The top panel of Figure~\ref{fig:shmr_0p3} compares the SHMR obtained using the fixed metal abundance of $0.3\,Z_\odot$ with the relation derived when adopting the {\sc Magneticum}-based metal abundance. At halo masses of $\log(M_{\rm 200}/M_\odot) > 13$, the SHMR is unchanged between the two assumptions. At lower halo masses, the metallicity inferred from the simulation leads to slightly higher stellar-to-halo mass ratios compared to the fixed-abundance case, although for the most bins the difference remains within the statistical uncertainties.}

{This behavior indicates that the lowest-mass bins, {corresponding to a regime closer to individual galaxies than to clusters,} are more sensitive to the assumed metal abundance in the spectral modeling. In contrast, the average temperatures of groups and clusters with $\log(M_{\rm 200}/M_\odot) \gtrsim 13$ are not that sensitive to the choice of metallicity, consistent with previous findings \citep{2025A&A...700A.167T}.} {For the lowest-mass bins, the predicted gas metallicity is around $0.6\,Z_\odot$, which allows us to test how the SHMR changes across a physically expected range of metallicities. Higher metallicity shifts the peak of the star formation efficiency to lower halo masses ($\sim 3\cdot10^{11}\,M_\odot$) and increases the stellar-to-halo mass ratios. However, this also alters the shape of the SHMR: with this peak position, the low-mass side before the peak no longer aligns with observational constraints from individual galaxies based on kinematics \citep{2025A&A...699A.311M, 2021ApJ...923...35M}, supporting our choice of $0.3\,Z_\odot$ as the fiducial abundance.}

\subsection{Comparison with simulations}

\begin{figure}
\centering
\includegraphics[width=1\hsize]{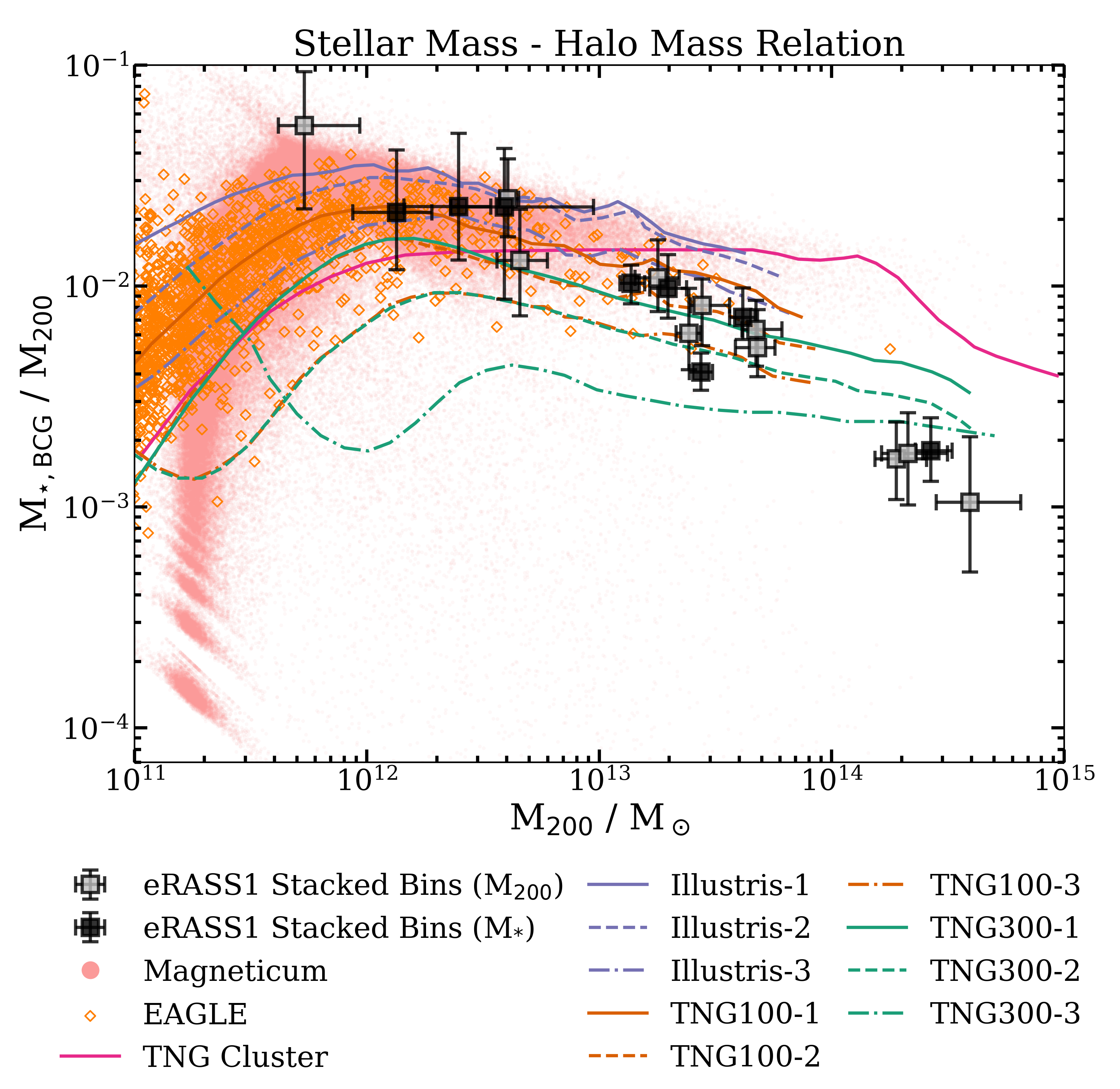}
  \caption{BCG stellar-to-halo mass ratio as a function of halo mass: comparison with simulations. Black points show stacked results in bins of stellar mass, while gray points show stacked results in bins of halo mass. Colored lines and symbols represent predictions from state-of-the-art hydrodynamical simulations.\label{fig:shmr_sim}}
\end{figure}

Figure~\ref{fig:shmr_sim} compares our SHMR measurements with predictions from state-of-the-art hydrodynamical simulations{: {\sc Magneticum} \citep{2016MNRAS.463.1797D, 2025arXiv250401061D}, EAGLE \citep{2015MNRAS.450.1937C, 2015MNRAS.446..521S}, Illustris \citep{2014MNRAS.444.1518V, 2014MNRAS.445..175G, 2015MNRAS.452..575S}, TNG-Cluster \citep{2024A&A...686A.157N, 2024A&A...686A..86R}, and different resolution runs of TNG300 and TNG100 \citep{2018MNRAS.480.5113M,2018MNRAS.475..648P,2018MNRAS.475..676S, 2018MNRAS.477.1206N, 2018MNRAS.475..624N}.} {For {\sc Magneticum}, we used the light cone covering an area of $30 \times 30\,\mathrm{deg}^2$ out to $z = 0.2$, sampling the local Universe. Galaxy clusters, groups, and their member galaxies were identified using the SubFind halo finder \citep{2001MNRAS.328..726S, 2009MNRAS.399..497D}. A detailed description of the light cone construction is provided in \citet{ilaria_opticallightcone}. The SHMRs from Illustris, TNG300, TNG100, and TNG-Cluster were obtained via the public data access of the TNG Project database\footnote{\url{https://www.tng-project.org/data/}}. For TNG we used snapshot~99, and for Illustris snapshot~135, both corresponding to $z = 0$. The SHMRs are based on precomputed values for central galaxies, where the stellar mass of the BCG  ($M_{\star,\mathrm{BCG}}$) is measured within twice the stellar half-mass radius. For EAGLE, we used data retrieved from the public EAGLE database\footnote{\url{https://eagle.strw.leidenuniv.nl/}}. We selected snapshot~27 ($z = 0.10$) from the reference model of the 50~Mpc medium-volume simulation described in \citet{2015MNRAS.446..521S}. In this case, the stellar mass of the central galaxy, $M_{\star,\mathrm{BCG}}$, is defined as the total mass of all star particles gravitationally bound to the galaxy. For all the simulations, the halo mass $M_{200}$ is defined as the total mass of the parent dark matter halo enclosed within $R_{200}$.}

{There are some major challenges in this kind of comparison. The first lies in the physical implementations in the simulations themselves. Feedback prescriptions, ICL formation efficiency, and stellar mass loss can all contribute to differences in the predicted SHMR among different simulations. Another} major challenge in this comparison arises from the different methods used to estimate stellar masses for BCGs. In observations, stellar mass is typically measured within an aperture defined by the galaxy’s surface brightness profile, often scaled to its effective radius. In simulations {we used in this work, the adopted definitions vary: stellar masses may be measured within the stellar half-mass radius or by summing all gravitationally bound star particles. These choices do not necessarily correspond to observational apertures and can lead to systematic offsets,} particularly at the high-mass end, where stars associated with the ICL may be partially or fully included in the BCG mass \citep[see][for such an attempt]{Cui_2014}. We believe {that these discrepancies can} contribute to the significant offsets seen in Fig.~\ref{fig:shmr_sim} between the simulations and our observational results.

We find broad agreement near the SHMR knee, albeit with substantial scatter, particularly for the {\sc Magneticum}, EAGLE, and Illustris simulations. {In the halo mass range above $10^{13}$ $M_{\odot}$ and below $10^{14}$ $M_{\odot}$,  EAGLE, TNG100, and TNG300-1 simulations show the closest agreement with the data, whereas Illustris and {\sc Magneticum} tend to slightly overpredict the stellar-to-halo mass ratio. The lower-resolution TNG300-2 and TNG300-3 runs underpredict the ratio across both mass ranges, highlighting the impact of resolution effects. The largest discrepancies occur at the highest halo masses, between $10^{14}$ and $10^{15}$ $M_{\odot}$. In this regime, only TNG-Cluster and TNG300 provide robust predictions, as simulations with smaller volumes{, or for which only a limited-size light-cone was used in this work,} do not sample such massive systems. Among these, TNG-Cluster shows the strongest offset, overpredicting the SHMR more significantly than TNG300, {which may reflect inaccuracies in the implemented sub-grid physics.}}

\section{Conclusions}

We have presented the SHMR over the widest halo mass range ($M_{\rm halo} \sim 10^{12}$--$10^{15}\ M_\odot$) ever probed, by using independent estimates of central galaxy stellar masses and host halo masses, derived from eROSITA eRASS1 X-ray data. Our analysis employs the spectral stacking technique described and validated in \cite{2025A&A...700A.167T} to trace the average X-ray spectra of optically selected groups from the \citet{yang_galaxy_2007} catalog. Each stacked spectrum was modeled to estimate the hot gas temperature, which was subsequently converted to halo mass via the \mbox{$M$--{$T_\mathrm{X}$}} relation of \cite{lovisari_relation}.

We find that the SHMR peaks near ${M_{200,\mathrm{X}}} \sim 10^{12}$$M_\odot$, followed by a decline in the stellar mass fraction toward higher halo masses. This behavior reflects the increasing impact of AGN feedback, which suppresses in situ star formation, while halo growth proceeds through dark matter accretion and mergers. At these higher masses, the central galaxy builds up its stellar content predominantly via dry mergers. {We note that the lowest-mass bins near the peak probe the low-energy sensitivity limit of eROSITA; deeper pointed observations with {\textit{XMM-Newton}} \citep{2001A&A...365L...1J} and future facilities such as NewAthena \citep{2025NatAs...9...36C} will be essential to robustly probe this mass regime.}

Our measurements are broadly consistent with previous observational studies based on WL, X-ray analyses of massive clusters, and satellite kinematics, as well as with semiempirical approaches such as abundance matching and empirical modeling. Comparisons with hydrodynamical simulations reveal discrepancies at the high-mass end, suggesting that current implementations of AGN feedback and star formation efficiency may require further refinement to reproduce the observed SHMR across the full halo mass range.

Finally, the SHMR is known to depend on assembly history and galaxy properties. Variations with BCG morphology, star formation rate, cluster concentration, and BCG-satellite luminosity contrast have been known from the previous studies \citep{2018MNRAS.477.1822M,2016MNRAS.457.3200M,2006MNRAS.368..715M,2015MNRAS.447..298H,2018ApJ...860....2G,2013MNRAS.428.2407W}. Future deep, wide-field X-ray surveys such as eRASS:4 will make it possible to extend this analysis to smaller, well-defined subsamples, enabling a more detailed exploration of these secondary dependencies.

\begin{acknowledgements}
{The authors thank the referee, Prof. I-Non Chiu, for helpful comments that improved the manuscript.} VT, PP, and IM acknowledge support from the European Research Council (ERC) under the European Union’s Horizon Europe research and innovation programme ERC CoG (Grant agreement No. 101045437, PI P. Popesso). KD acknowledges support by the COMPLEX project from the European Research Council (ERC) under the European Union’s Horizon 2020 research and innovation program grant agreement ERC-2019-AdG 882679. The calculations for the Magneticum simulations were carried out at the Leibniz Supercomputer Center (LRZ) under the project pr83li. SVZ and VB acknowledge support by the \emph{Deut\-sche For\-schungs\-ge\-mein\-schaft, DFG\/} project nr. 415510302. LL acknowledges support from INAF grant 1.05.12.04.01. SE acknowledges financial contribution from the contracts Prin-MUR 2022 supported by Next Generation EU (M4.C2.1.1, n.20227RNLY3 {\it The concordance cosmological model: stress-tests with galaxy clusters}), and from the European Union’s Horizon 2020 Programme under the AHEAD2020 project (grant agreement n. 871158).  This research has made using software provided by the Chandra X-ray Center (CXC) in the application packages CIAO and Sherpa. This work is based on data from eROSITA, the soft X-ray instrument aboard SRG, a joint Russian-German science mission supported by the Russian Space Agency (Roskosmos), in the interests of the Russian Academy of Sciences represented by its Space Research Institute (IKI), and the Deutsches Zentrum für Luft- und Raumfahrt (DLR). The SRG spacecraft was built by Lavochkin Association (NPOL) and its subcontractors, and is operated by NPOL with support from the Max Planck Institute for Extraterrestrial Physics (MPE). The development and construction of the eROSITA X-ray instrument was led by MPE, with contributions from the Dr. Karl Remeis Observatory Bamberg \& ECAP (FAU Erlangen-Nuernberg), the University of Hamburg Observatory, the Leibniz Institute for Astrophysics Potsdam (AIP), and the Institute for Astronomy and Astrophysics of the University of Tübingen, with the support of DLR and the Max Planck Society. The Argelander Institute for Astronomy of the University of Bonn and the Ludwig Maximilians Universität Munich also participated in the science preparation for eROSITA. The eROSITA data shown here were processed using the eSASS software system developed by the German eROSITA consortium. We acknowledge the Virgo Consortium for making their simulation data available. The EAGLE simulations were performed using the DiRAC-2 facility at Durham, managed by the ICC, and the PRACE facility Curie based in France at TGCC, CEA, Bruy\`eres-le-Ch\^atel.
\end{acknowledgements}

  \bibliographystyle{aa.bst} 
  \bibliography{paper.bib} 

@ARTICLE{Cui_2014,
       author = {{Cui}, Weiguang and {Murante}, G. and {Monaco}, P. and {Borgani}, S. and {Granato}, G.~L. and {Killedar}, M. and {De Lucia}, G. and {Presotto}, V. and {Dolag}, K.},
        title = "{Characterizing diffused stellar light in simulated galaxy clusters}",
      journal = {\mnras},
         year = 2014,
        month = jan,
       volume = {437},
       number = {1},
        pages = {816-830},
          doi = {10.1093/mnras/stt1940},
archivePrefix = {arXiv},
       eprint = {1310.1396},
 primaryClass = {astro-ph.CO},
       adsurl = {https://ui.adsabs.harvard.edu/abs/2014MNRAS.437..816C}
}

@ARTICLE{Yang_2009,
       author = {{Yang}, Xiaohu and {Mo}, H.~J. and {van den Bosch}, Frank C.},
        title = "{Galaxy Groups in the SDSS DR4. III. The Luminosity and Stellar Mass Functions}",
      journal = {\apj},
         year = 2009,
        month = apr,
       volume = {695},
       number = {2},
        pages = {900-916},
          doi = {10.1088/0004-637X/695/2/900},
archivePrefix = {arXiv},
       eprint = {0808.0539},
 primaryClass = {astro-ph},
       adsurl = {https://ui.adsabs.harvard.edu/abs/2009ApJ...695..900Y}
}

@ARTICLE{Zheng_2023,
       author = {{Zheng}, Yun-Liang and {Yang}, Xiaohu and {He}, Min and {Shen}, Shi-Yin and {Li}, Qingyang and {Li}, Xuejie},
        title = "{Measuring the X-ray luminosities of DESI groups from eROSITA Final Equatorial-Depth Survey - I. X-ray luminosity-halo mass scaling relation}",
      journal = {\mnras},
         year = 2023,
        month = aug,
       volume = {523},
       number = {4},
        pages = {4909-4922},
          doi = {10.1093/mnras/stad1684},
archivePrefix = {arXiv},
       eprint = {2306.02594},
 primaryClass = {astro-ph.GA},
       adsurl = {https://ui.adsabs.harvard.edu/abs/2023MNRAS.523.4909Z}
}

@ARTICLE{Wang_2022,
       author = {{Wang}, Jiaqi and {Yang}, Xiaohu and {Zhang}, Jun and {Li}, Hekun and {Fong}, Matthew and {Xu}, Haojie and {He}, Min and {Gu}, Yizhou and {Luo}, Wentao and {Dong}, Fuyu and {Wang}, Yirong and {Li}, Qingyang and {Katsianis}, Antonios and {Wang}, Haoran and {Shen}, Zhi and {Alonso Vaquero}, Pedro and {Liu}, Cong and {Huang}, Yiqi and {Liu}, Zhenjie},
        title = "{Halo Properties and Mass Functions of Groups/Clusters from the DESI Legacy Imaging Surveys DR9}",
      journal = {\apj},
         year = 2022,
        month = sep,
       volume = {936},
       number = {2},
          eid = {161},
        pages = {161},
          doi = {10.3847/1538-4357/ac8986},
archivePrefix = {arXiv},
       eprint = {2207.12771},
 primaryClass = {astro-ph.CO},
       adsurl = {https://ui.adsabs.harvard.edu/abs/2022ApJ...936..161W}
}

@article{yang_galaxy_2007,
	title = {Galaxy {Groups} in the {SDSS} {DR4}. {I}. {The} {Catalog} and {Basic} {Properties}},
	volume = {671},
	issn = {0004-637X},
	url = {https://ui.adsabs.harvard.edu/abs/2007ApJ...671..153Y},
	doi = {10.1086/522027},
	urldate = {2023-08-24},
	journal = {The Astrophysical Journal},
	author = {Yang, Xiaohu and Mo, H. J. and van den Bosch, Frank C. and Pasquali, Anna and Li, Cheng and Barden, Marco},
	month = dec,
	year = {2007},
	note = {ADS Bibcode: 2007ApJ...671..153Y},
	pages = {153--170},
}

@ARTICLE{ilaria_lightcone,
       author = {{Marini}, I. and {Popesso}, P. and {Lamer}, G. and {Dolag}, K. and {Biffi}, V. and {Vladutescu-Zopp}, S. and {Dev}, A. and {Toptun}, V. and {Bulbul}, E. and {Comparat}, J. and {Malavasi}, N. and {Merloni}, A. and {Mroczkowski}, T. and {Ponti}, G. and {Seppi}, R. and {Shreeram}, S. and {Zhang}, Y.},
        title = "{Detecting galaxy groups populating the local Universe in the eROSITA era}",
      journal = {\aap},
         year = 2024,
        month = sep,
       volume = {689},
          eid = {A7},
        pages = {A7},
          doi = {10.1051/0004-6361/202450442},
archivePrefix = {arXiv},
       eprint = {2404.12719},
 primaryClass = {astro-ph.CO},
       adsurl = {https://ui.adsabs.harvard.edu/abs/2024A&A...689A...7M}
}

@software{sherpa,
  author       = {Doug Burke and
                  Omar Laurino and
                  wmclaugh and
                  Hans Moritz Günther and
                  Marie-Terrell and
                  dtnguyen2 and
                  Aneta Siemiginowska and
                  Harlan Cheer and
                  Jamie Budynkiewicz and
                  Tom Aldcroft and
                  Christoph Deil and
                  Brigitta Sipőcz and
                  Johannes Buchner and
                  nplee and
                  Axel Donath and
                  Iva Laginja and
                  Katrin Leinweber and
                  Todd},
  title        = {sherpa/sherpa: Sherpa 4.16.0},
  month        = oct,
  year         = 2023,
  publisher    = {Zenodo},
  version      = {4.16.0},
  doi          = {10.5281/zenodo.825839},
  url          = {https://doi.org/10.5281/zenodo.825839}
}

@INPROCEEDINGS{abundances_1,
       author = {{Baumgartner}, W.~H. and {Loewenstein}, M. and {Horner}, D.~J. and {Mushotzky}, R.~F.},
        title = "{Intermediate Element Abundances in Galaxy Clusters}",
    booktitle = {AAS/High Energy Astrophysics Division \#7},
         year = 2003,
       series = {AAS/High Energy Astrophysics Division},
       volume = {7},
        month = mar,
          eid = {35.03},
        pages = {35.03},
       adsurl = {https://ui.adsabs.harvard.edu/abs/2003HEAD....7.3503B}
}

@ARTICLE{gastadello_review,
       author = {{Gastaldello}, Fabio and {Simionescu}, Aurora and {Mernier}, Francois and {Biffi}, Veronica and {Gaspari}, Massimo and {Sato}, Kosuke and {Matsushita}, Kyoko},
        title = "{The Metal Content of the Hot Atmospheres of Galaxy Groups}",
      journal = {Universe},
         year = 2021,
        month = jun,
       volume = {7},
       number = {7},
          eid = {208},
        pages = {208},
          doi = {10.3390/universe7070208},
archivePrefix = {arXiv},
       eprint = {2106.13258},
 primaryClass = {astro-ph.CO},
       adsurl = {https://ui.adsabs.harvard.edu/abs/2021Univ....7..208G}
}

@ARTICLE{abundances_xcop,
       author = {{Ghizzardi}, Simona and {Molendi}, Silvano and {van der Burg}, Remco and {De Grandi}, Sabrina and {Bartalucci}, Iacopo and {Gastaldello}, Fabio and {Rossetti}, Mariachiara and {Biffi}, Veronica and {Borgani}, Stefano and {Eckert}, Dominique and {Ettori}, Stefano and {Gaspari}, Massimo and {Ghirardini}, Vittorio and {Rasia}, Elena},
        title = "{Iron in X-COP: Tracing enrichment in cluster outskirts with high accuracy abundance profiles}",
      journal = {\aap},
         year = 2021,
        month = feb,
       volume = {646},
          eid = {A92},
        pages = {A92},
          doi = {10.1051/0004-6361/202038501},
archivePrefix = {arXiv},
       eprint = {2007.01084},
 primaryClass = {astro-ph.CO},
       adsurl = {https://ui.adsabs.harvard.edu/abs/2021A&A...646A..92G}
}

@ARTICLE{abundances_suzaku,
       author = {{Urban}, O. and {Werner}, N. and {Allen}, S.~W. and {Simionescu}, A. and {Mantz}, A.},
        title = "{A uniform metallicity in the outskirts of massive, nearby galaxy clusters}",
      journal = {\mnras},
         year = 2017,
        month = oct,
       volume = {470},
       number = {4},
        pages = {4583-4599},
          doi = {10.1093/mnras/stx1542},
archivePrefix = {arXiv},
       eprint = {1706.01567},
 primaryClass = {astro-ph.CO},
       adsurl = {https://ui.adsabs.harvard.edu/abs/2017MNRAS.470.4583U}
}

@ARTICLE{lovisari_relation,
       author = {{Lovisari}, L. and {Reiprich}, T.~H. and {Schellenberger}, G.},
        title = "{Scaling properties of a complete X-ray selected galaxy group sample}",
      journal = {\aap},
         year = 2015,
        month = jan,
       volume = {573},
          eid = {A118},
        pages = {A118},
          doi = {10.1051/0004-6361/201423954},
archivePrefix = {arXiv},
       eprint = {1409.3845},
 primaryClass = {astro-ph.CO},
       adsurl = {https://ui.adsabs.harvard.edu/abs/2015A&A...573A.118L}
}

@ARTICLE{robotham,
       author = {{Robotham}, A.~S.~G. and {Norberg}, P. and {Driver}, S.~P. and {Baldry}, I.~K. and {Bamford}, S.~P. and {Hopkins}, A.~M. and {Liske}, J. and {Loveday}, J. and {Merson}, A. and {Peacock}, J.~A. and {Brough}, S. and {Cameron}, E. and {Conselice}, C.~J. and {Croom}, S.~M. and {Frenk}, C.~S. and {Gunawardhana}, M. and {Hill}, D.~T. and {Jones}, D.~H. and {Kelvin}, L.~S. and {Kuijken}, K. and {Nichol}, R.~C. and {Parkinson}, H.~R. and {Pimbblet}, K.~A. and {Phillipps}, S. and {Popescu}, C.~C. and {Prescott}, M. and {Sharp}, R.~G. and {Sutherland}, W.~J. and {Taylor}, E.~N. and {Thomas}, D. and {Tuffs}, R.~J. and {van Kampen}, E. and {Wijesinghe}, D.},
        title = "{Galaxy and Mass Assembly (GAMA): the GAMA galaxy group catalogue (G$^{3}$Cv1)}",
      journal = {\mnras},
         year = 2011,
        month = oct,
       volume = {416},
       number = {4},
        pages = {2640-2668},
          doi = {10.1111/j.1365-2966.2011.19217.x},
archivePrefix = {arXiv},
       eprint = {1106.1994},
 primaryClass = {astro-ph.CO},
       adsurl = {https://ui.adsabs.harvard.edu/abs/2011MNRAS.416.2640R}
}

@ARTICLE{paola_24,
       author = {{Popesso}, P. and {Biviano}, A. and {Bulbul}, E. and {Merloni}, A. and {Comparat}, J. and {Clerc}, N. and {Igo}, Z. and {Liu}, A. and {Driver}, S. and {Salvato}, M. and {Brusa}, M. and {Bahar}, Y.~E. and {Malavasi}, N. and {Ghirardini}, V. and {Robotham}, A. and {Liske}, J. and {Grandis}, S.},
        title = "{The X-ray invisible Universe. A look into the haloes undetected by eROSITA}",
      journal = {\mnras},
         year = 2024,
        month = jan,
       volume = {527},
       number = {1},
        pages = {895-910},
          doi = {10.1093/mnras/stad3253},
archivePrefix = {arXiv},
       eprint = {2302.08405},
 primaryClass = {astro-ph.CO},
       adsurl = {https://ui.adsabs.harvard.edu/abs/2024MNRAS.527..895P}
}

@ARTICLE{2024A&A...685A.106B,
       author = {{Bulbul}, E. and {Liu}, A. and {Kluge}, M. and {Zhang}, X. and {Sanders}, J.~S. and {Bahar}, Y.~E. and {Ghirardini}, V. and {Artis}, E. and {Seppi}, R. and {Garrel}, C. and {Ramos-Ceja}, M.~E. and {Comparat}, J. and {Balzer}, F. and {B{\"o}ckmann}, K. and {Br{\"u}ggen}, M. and {Clerc}, N. and {Dennerl}, K. and {Dolag}, K. and {Freyberg}, M. and {Grandis}, S. and {Gruen}, D. and {Kleinebreil}, F. and {Krippendorf}, S. and {Lamer}, G. and {Merloni}, A. and {Migkas}, K. and {Nandra}, K. and {Pacaud}, F. and {Predehl}, P. and {Reiprich}, T.~H. and {Schrabback}, T. and {Veronica}, A. and {Weller}, J. and {Zelmer}, S.},
        title = "{The SRG/eROSITA All-Sky Survey. The first catalog of galaxy clusters and groups in the Western Galactic Hemisphere}",
      journal = {\aap},
         year = 2024,
        month = may,
       volume = {685},
          eid = {A106},
        pages = {A106},
          doi = {10.1051/0004-6361/202348264},
archivePrefix = {arXiv},
       eprint = {2402.08452},
 primaryClass = {astro-ph.CO},
       adsurl = {https://ui.adsabs.harvard.edu/abs/2024A&A...685A.106B}
}

@ARTICLE{eagle,
       author = {{Barnes}, David J. and {Kay}, Scott T. and {Bah{\'e}}, Yannick M. and {Dalla Vecchia}, Claudio and {McCarthy}, Ian G. and {Schaye}, Joop and {Bower}, Richard G. and {Jenkins}, Adrian and {Thomas}, Peter A. and {Schaller}, Matthieu and {Crain}, Robert A. and {Theuns}, Tom and {White}, Simon D.~M.},
        title = "{The Cluster-EAGLE project: global properties of simulated clusters with resolved galaxies}",
      journal = {\mnras},
         year = 2017,
        month = oct,
       volume = {471},
       number = {1},
        pages = {1088-1106},
          doi = {10.1093/mnras/stx1647},
archivePrefix = {arXiv},
       eprint = {1703.10907},
 primaryClass = {astro-ph.GA},
       adsurl = {https://ui.adsabs.harvard.edu/abs/2017MNRAS.471.1088B}
}

@ARTICLE{2009A&A...498..361P,
       author = {{Pratt}, G.~W. and {Croston}, J.~H. and {Arnaud}, M. and {B{\"o}hringer}, H.},
        title = "{Galaxy cluster X-ray luminosity scaling relations from a representative local sample (REXCESS)}",
      journal = {\aap},
         year = 2009,
        month = may,
       volume = {498},
       number = {2},
        pages = {361-378},
          doi = {10.1051/0004-6361/200810994},
archivePrefix = {arXiv},
       eprint = {0809.3784},
 primaryClass = {astro-ph},
       adsurl = {https://ui.adsabs.harvard.edu/abs/2009A&A...498..361P}
}

@ARTICLE{ilaria_opticallightcone,
       author = {{Marini}, I. and {Popesso}, P. and {Dolag}, K. and {Bravo}, M. and {Robotham}, A. and {Tempel}, E. and {Li}, Q. and {Yang}, X. and {Csizi}, B. and {Behroozi}, P. and {Biffi}, V. and {Biviano}, A. and {Lamer}, G. and {Malavasi}, N. and {Mazengo}, D. and {Toptun}, V.},
        title = "{Detecting clusters and groups of galaxies populating the local Universe in large optical spectroscopic surveys}",
      journal = {\aap},
         year = 2025,
        month = feb,
       volume = {694},
          eid = {A207},
        pages = {A207},
          doi = {10.1051/0004-6361/202452028},
archivePrefix = {arXiv},
       eprint = {2411.16455},
 primaryClass = {astro-ph.GA},
       adsurl = {https://ui.adsabs.harvard.edu/abs/2025A&A...694A.207M}
}

@ARTICLE{paola_stacking_magneticum,
       author = {{Popesso}, P. and {Marini}, I. and {Dolag}, K. and {Lamer}, G. and {Csizi}, B. and {Vladutescu-Zopp}, S. and {Biffi}, V. and {Robothan}, A. and {Bravo}, M. and {Tempel}, E. and {Yang}, X. and {Li}, Q. and {Biviano}, A. and {Lovisari}, L. and {Ettori}, S. and {Angelinelli}, M. and {Driver}, S. and {Toptun}, V. and {Dev}, A. and {Mazengo}, D. and {Merloni}, A. and {Mroczkowski}, T. and {Comparat}, J. and {Zhang}, Y. and {Ponti}, G. and {Bulbul}, E.},
        title = "{The perils of stacking optically selected groups in eROSITA data: The Magneticum perspective}",
      journal = {\aap},
         year = 2025,
        month = dec,
       volume = {704},
          eid = {A277},
        pages = {A277},
          doi = {10.1051/0004-6361/202453253},
archivePrefix = {arXiv},
       eprint = {2411.16546},
 primaryClass = {astro-ph.GA},
       adsurl = {https://ui.adsabs.harvard.edu/abs/2025A&A...704A.277P}
}

@ARTICLE{paola_gasfraction,
       author = {{Popesso}, P. and {Biviano}, A. and {Marini}, I. and {Dolag}, K. and {Vladutescu-Zopp}, S. and {Csizi}, B. and {Biffi}, V. and {Lamer}, G. and {Robothan}, A. and {Bravo}, M. and {Lovisari}, L. and {Ettori}, S. and {Angelinelli}, M. and {Driver}, S. and {Toptun}, V. and {Dev}, A. and {Mazengo}, D. and {Merloni}, A. and {Comparat}, J. and {Ponti}, G. and {Mroczkowski}, T. and {Bulbul}, E. and {Grandis}, S. and {Bahar}, E.},
        title = "{The hot gas mass fraction in halos: From Milky Way-like groups to massive clusters}",
      journal = {\aap},
         year = 2026,
        month = mar,
       volume = {707},
          eid = {A362},
        pages = {A362},
          doi = {10.1051/0004-6361/202453256},
archivePrefix = {arXiv},
       eprint = {2411.16555},
 primaryClass = {astro-ph.GA},
       adsurl = {https://ui.adsabs.harvard.edu/abs/2026A&A...707A.362P}
}

@ARTICLE{paola_profiles,
       author = {{Popesso}, P. and {Marini}, I. and {Dolag}, K. and {Lamer}, G. and {Csizi}, B. and {Biffi}, V. and {Robothan}, A. and {Bravo}, M. and {Biviano}, A. and {Vladutescu-Zopp}, S. and {Lovisari}, L. and {Ettori}, S. and {Angelinelli}, M. and {Driver}, S. and {Toptun}, V. and {Dev}, A. and {Mazengo}, D. and {Merloni}, A. and {Zhang}, Y. and {Comparat}, J. and {Ponti}, G. and {Mroczkowski}, T. and {Bulbul}, E.},
        title = "{Average X-ray properties of galaxy groups: From Milky Way-like halos to massive clusters}",
      journal = {\aap},
         year = 2025,
        month = dec,
       volume = {704},
          eid = {A278},
        pages = {A278},
          doi = {10.1051/0004-6361/202453255},
archivePrefix = {arXiv},
       eprint = {2411.17120},
 primaryClass = {astro-ph.GA},
       adsurl = {https://ui.adsabs.harvard.edu/abs/2025A&A...704A.278P}
}

@ARTICLE{2001MNRAS.328..726S,
       author = {{Springel}, Volker and {White}, Simon D.~M. and {Tormen}, Giuseppe and {Kauffmann}, Guinevere},
        title = "{Populating a cluster of galaxies - I. Results at z=0}",
      journal = {\mnras},
         year = 2001,
        month = dec,
       volume = {328},
       number = {3},
        pages = {726-750},
          doi = {10.1046/j.1365-8711.2001.04912.x},
archivePrefix = {arXiv},
       eprint = {astro-ph/0012055},
 primaryClass = {astro-ph},
       adsurl = {https://ui.adsabs.harvard.edu/abs/2001MNRAS.328..726S}
}

@ARTICLE{2009MNRAS.399..497D,
       author = {{Dolag}, K. and {Borgani}, S. and {Murante}, G. and {Springel}, V.},
        title = "{Substructures in hydrodynamical cluster simulations}",
      journal = {\mnras},
         year = 2009,
        month = oct,
       volume = {399},
       number = {2},
        pages = {497-514},
          doi = {10.1111/j.1365-2966.2009.15034.x},
archivePrefix = {arXiv},
       eprint = {0808.3401},
 primaryClass = {astro-ph},
       adsurl = {https://ui.adsabs.harvard.edu/abs/2009MNRAS.399..497D}
}

@ARTICLE{2024A&A...682A..34M,
       author = {{Merloni}, A. and {Lamer}, G. and {Liu}, T. and {Ramos-Ceja}, M.~E. and {Brunner}, H. and {Bulbul}, E. and {Dennerl}, K. and {Doroshenko}, V. and {Freyberg}, M.~J. and {Friedrich}, S. and {Gatuzz}, E. and {Georgakakis}, A. and {Haberl}, F. and {Igo}, Z. and {Kreykenbohm}, I. and {Liu}, A. and {Maitra}, C. and {Malyali}, A. and {Mayer}, M.~G.~F. and {Nandra}, K. and {Predehl}, P. and {Robrade}, J. and {Salvato}, M. and {Sanders}, J.~S. and {Stewart}, I. and {Tub{\'\i}n-Arenas}, D. and {Weber}, P. and {Wilms}, J. and {Arcodia}, R. and {Artis}, E. and {Aschersleben}, J. and {Avakyan}, A. and {Aydar}, C. and {Bahar}, Y.~E. and {Balzer}, F. and {Becker}, W. and {Berger}, K. and {Boller}, T. and {Bornemann}, W. and {Br{\"u}ggen}, M. and {Brusa}, M. and {Buchner}, J. and {Burwitz}, V. and {Camilloni}, F. and {Clerc}, N. and {Comparat}, J. and {Coutinho}, D. and {Czesla}, S. and {Dannhauer}, S.~M. and {Dauner}, L. and {Dauser}, T. and {Dietl}, J. and {Dolag}, K. and {Dwelly}, T. and {Egg}, K. and {Ehl}, E. and {Freund}, S. and {Friedrich}, P. and {Gaida}, R. and {Garrel}, C. and {Ghirardini}, V. and {Gokus}, A. and {Gr{\"u}nwald}, G. and {Grandis}, S. and {Grotova}, I. and {Gruen}, D. and {Gueguen}, A. and {H{\"a}mmerich}, S. and {Hamaus}, N. and {Hasinger}, G. and {Haubner}, K. and {Homan}, D. and {Ider Chitham}, J. and {Joseph}, W.~M. and {Joyce}, A. and {K{\"o}nig}, O. and {Kaltenbrunner}, D.~M. and {Khokhriakova}, A. and {Kink}, W. and {Kirsch}, C. and {Kluge}, M. and {Knies}, J. and {Krippendorf}, S. and {Krumpe}, M. and {Kurpas}, J. and {Li}, P. and {Liu}, Z. and {Locatelli}, N. and {Lorenz}, M. and {M{\"u}ller}, S. and {Magaudda}, E. and {Mannes}, C. and {McCall}, H. and {Meidinger}, N. and {Michailidis}, M. and {Migkas}, K. and {Mu{\~n}oz-Giraldo}, D. and {Musiimenta}, B. and {Nguyen-Dang}, N.~T. and {Ni}, Q. and {Olechowska}, A. and {Ota}, N. and {Pacaud}, F. and {Pasini}, T. and {Perinati}, E. and {Pires}, A.~M. and {Pommranz}, C. and {Ponti}, G. and {Poppenhaeger}, K. and {P{\"u}hlhofer}, G. and {Rau}, A. and {Reh}, M. and {Reiprich}, T.~H. and {Roster}, W. and {Saeedi}, S. and {Santangelo}, A. and {Sasaki}, M. and {Schmitt}, J. and {Schneider}, P.~C. and {Schrabback}, T. and {Schuster}, N. and {Schwope}, A. and {Seppi}, R. and {Serim}, M.~M. and {Shreeram}, S. and {Sokolova-Lapa}, E. and {Starck}, H. and {Stelzer}, B. and {Stierhof}, J. and {Suleimanov}, V. and {Tenzer}, C. and {Traulsen}, I. and {Tr{\"u}mper}, J. and {Tsuge}, K. and {Urrutia}, T. and {Veronica}, A. and {Waddell}, S.~G.~H. and {Willer}, R. and {Wolf}, J. and {Yeung}, M.~C.~H. and {Zainab}, A. and {Zangrandi}, F. and {Zhang}, X. and {Zhang}, Y. and {Zheng}, X.},
        title = "{The SRG/eROSITA all-sky survey. First X-ray catalogues and data release of the western Galactic hemisphere}",
      journal = {\aap},
         year = 2024,
        month = feb,
       volume = {682},
          eid = {A34},
        pages = {A34},
          doi = {10.1051/0004-6361/202347165},
archivePrefix = {arXiv},
       eprint = {2401.17274},
 primaryClass = {astro-ph.HE},
       adsurl = {https://ui.adsabs.harvard.edu/abs/2024A&A...682A..34M}
}

@ARTICLE{1989GeCoA..53..197A,
       author = {{Anders}, E. and {Grevesse}, N.},
        title = "{Abundances of the elements: Meteoritic and solar}",
      journal = {\gca},
         year = 1989,
        month = jan,
       volume = {53},
       number = {1},
        pages = {197-214},
          doi = {10.1016/0016-7037(89)90286-X},
       adsurl = {https://ui.adsabs.harvard.edu/abs/1989GeCoA..53..197A}
}

@ARTICLE{1985A&AS...62..197M,
       author = {{Mewe}, R. and {Gronenschild}, E.~H.~B.~M. and {van den Oord}, G.~H.~J.},
        title = "{Calculated X-Radiation from Optically Thin Plasmas - Part Five}",
      journal = {\aaps},
         year = 1985,
        month = nov,
       volume = {62},
        pages = {197},
       adsurl = {https://ui.adsabs.harvard.edu/abs/1985A&AS...62..197M}
}

@ARTICLE{1986A&AS...65..511M,
       author = {{Mewe}, R. and {Lemen}, J.~R. and {van den Oord}, G.~H.~J.},
        title = "{Calculated X-radiation from optically thin plasmas. VI - Improved calculations for continuum emission and approximation formulae for nonrelativistic average Gaunt actors.}",
      journal = {\aaps},
         year = 1986,
        month = sep,
       volume = {65},
        pages = {511-536},
       adsurl = {https://ui.adsabs.harvard.edu/abs/1986A&AS...65..511M}
}

@ARTICLE{1995ApJ...438L.115L,
       author = {{Liedahl}, Duane A. and {Osterheld}, Albert L. and {Goldstein}, William H.},
        title = "{New Calculations of Fe L-Shell X-Ray Spectra in High-Temperature Plasmas}",
      journal = {\apjl},
         year = 1995,
        month = jan,
       volume = {438},
        pages = {L115},
          doi = {10.1086/187729},
       adsurl = {https://ui.adsabs.harvard.edu/abs/1995ApJ...438L.115L}
}

@ARTICLE{2018ApJS..239...35D,
       author = {{Diemer}, Benedikt},
        title = "{COLOSSUS: A Python Toolkit for Cosmology, Large-scale Structure, and Dark Matter Halos}",
      journal = {\apjs},
         year = 2018,
        month = dec,
       volume = {239},
       number = {2},
          eid = {35},
        pages = {35},
          doi = {10.3847/1538-4365/aaee8c},
archivePrefix = {arXiv},
       eprint = {1712.04512},
 primaryClass = {astro-ph.CO},
       adsurl = {https://ui.adsabs.harvard.edu/abs/2018ApJS..239...35D}
}

@ARTICLE{2025arXiv250401061D,
       author = {{Dolag}, Klaus and {Remus}, Rhea-Silvia and {Valenzuela}, Lucas M. and {Kimmig}, Lucas C. and {Seidel}, Benjamin and {Fortune}, Silvio and {Stoiber}, Johannes and {Ivleva}, Anna and {Hoffmann}, Tadziu and {Biffi}, Veronica and {Marini}, Ilaria and {Popesso}, Paola and {Vladutescu-Zopp}, Stephan},
        title = "{Encyclopedia Magneticum: Scaling Relations from Cosmic Dawn to Present Day}",
      journal = {\aap},
         year = 2025,
        month = apr,
          eid = {arXiv:2504.01061},
        pages = {submitted},
          doi = {10.48550/arXiv.2504.01061},
archivePrefix = {arXiv},
       eprint = {2504.01061},
 primaryClass = {astro-ph.CO},
       adsurl = {https://ui.adsabs.harvard.edu/abs/2025arXiv250401061D}
}

@ARTICLE{2025A&A...700A.167T,
       author = {{Toptun}, V. and {Popesso}, P. and {Marini}, I. and {Dolag}, K. and {Lamer}, G. and {Yang}, X. and {Li}, Q. and {Csizi}, B. and {Lovisari}, L. and {Ettori}, S. and {Biffi}, V. and {Vladutescu-Zopp}, S. and {Dev}, A. and {Mazengo}, D. and {Merloni}, A. and {Comparat}, J. and {Ponti}, G. and {Bulbul}, E.},
        title = "{The eROSITA view on the halo mass─temperature relation: From low-mass groups to massive clusters}",
      journal = {\aap},
         year = 2025,
        month = aug,
       volume = {700},
          eid = {A167},
        pages = {A167},
          doi = {10.1051/0004-6361/202554352},
archivePrefix = {arXiv},
       eprint = {2505.01502},
 primaryClass = {astro-ph.GA},
       adsurl = {https://ui.adsabs.harvard.edu/abs/2025A&A...700A.167T}
}

@ARTICLE{2009ApJ...696..620C,
       author = {{Conroy}, Charlie and {Wechsler}, Risa H.},
        title = "{Connecting Galaxies, Halos, and Star Formation Rates Across Cosmic Time}",
      journal = {\apj},
         year = 2009,
        month = may,
       volume = {696},
       number = {1},
        pages = {620-635},
          doi = {10.1088/0004-637X/696/1/620},
archivePrefix = {arXiv},
       eprint = {0805.3346},
 primaryClass = {astro-ph},
       adsurl = {https://ui.adsabs.harvard.edu/abs/2009ApJ...696..620C}
}

@ARTICLE{2010ApJ...717..379B,
       author = {{Behroozi}, Peter S. and {Conroy}, Charlie and {Wechsler}, Risa H.},
        title = "{A Comprehensive Analysis of Uncertainties Affecting the Stellar Mass-Halo Mass Relation for 0 < z < 4}",
      journal = {\apj},
         year = 2010,
        month = jul,
       volume = {717},
       number = {1},
        pages = {379-403},
          doi = {10.1088/0004-637X/717/1/379},
archivePrefix = {arXiv},
       eprint = {1001.0015},
 primaryClass = {astro-ph.CO},
       adsurl = {https://ui.adsabs.harvard.edu/abs/2010ApJ...717..379B}
}

@ARTICLE{2010ApJ...710..903M,
       author = {{Moster}, Benjamin P. and {Somerville}, Rachel S. and {Maulbetsch}, Christian and {van den Bosch}, Frank C. and {Macci{\`o}}, Andrea V. and {Naab}, Thorsten and {Oser}, Ludwig},
        title = "{Constraints on the Relationship between Stellar Mass and Halo Mass at Low and High Redshift}",
      journal = {\apj},
         year = 2010,
        month = feb,
       volume = {710},
       number = {2},
        pages = {903-923},
          doi = {10.1088/0004-637X/710/2/903},
archivePrefix = {arXiv},
       eprint = {0903.4682},
 primaryClass = {astro-ph.CO},
       adsurl = {https://ui.adsabs.harvard.edu/abs/2010ApJ...710..903M}
}

@ARTICLE{2013MNRAS.428.3121M,
       author = {{Moster}, Benjamin P. and {Naab}, Thorsten and {White}, Simon D.~M.},
        title = "{Galactic star formation and accretion histories from matching galaxies to dark matter haloes}",
      journal = {\mnras},
         year = 2013,
        month = feb,
       volume = {428},
       number = {4},
        pages = {3121-3138},
          doi = {10.1093/mnras/sts261},
archivePrefix = {arXiv},
       eprint = {1205.5807},
 primaryClass = {astro-ph.CO},
       adsurl = {https://ui.adsabs.harvard.edu/abs/2013MNRAS.428.3121M}
}

@ARTICLE{2018MNRAS.477.1822M,
       author = {{Moster}, Benjamin P. and {Naab}, Thorsten and {White}, Simon D.~M.},
        title = "{EMERGE - an empirical model for the formation of galaxies since z {\ensuremath{\sim}} 10}",
      journal = {\mnras},
         year = 2018,
        month = jun,
       volume = {477},
       number = {2},
        pages = {1822-1852},
          doi = {10.1093/mnras/sty655},
archivePrefix = {arXiv},
       eprint = {1705.05373},
 primaryClass = {astro-ph.GA},
       adsurl = {https://ui.adsabs.harvard.edu/abs/2018MNRAS.477.1822M}
}

@ARTICLE{2017MNRAS.470..651R,
       author = {{Rodr{\'\i}guez-Puebla}, Aldo and {Primack}, Joel R. and {Avila-Reese}, Vladimir and {Faber}, S.~M.},
        title = "{Constraining the galaxy-halo connection over the last 13.3 Gyr: star formation histories, galaxy mergers and structural properties}",
      journal = {\mnras},
         year = 2017,
        month = sep,
       volume = {470},
       number = {1},
        pages = {651-687},
          doi = {10.1093/mnras/stx1172},
archivePrefix = {arXiv},
       eprint = {1703.04542},
 primaryClass = {astro-ph.GA},
       adsurl = {https://ui.adsabs.harvard.edu/abs/2017MNRAS.470..651R}
}

@ARTICLE{2015MNRAS.450.1604L,
       author = {{Lu}, Zhankui and {Mo}, H.~J. and {Lu}, Yu and {Katz}, Neal and {Weinberg}, Martin D. and {van den Bosch}, Frank C. and {Yang}, Xiaohu},
        title = "{Star formation and stellar mass assembly in dark matter haloes: from giants to dwarfs}",
      journal = {\mnras},
         year = 2015,
        month = jun,
       volume = {450},
       number = {2},
        pages = {1604-1617},
          doi = {10.1093/mnras/stv667},
archivePrefix = {arXiv},
       eprint = {1406.5068},
 primaryClass = {astro-ph.GA},
       adsurl = {https://ui.adsabs.harvard.edu/abs/2015MNRAS.450.1604L}
}

@ARTICLE{2013ApJ...777L..10B,
       author = {{Behroozi}, Peter S. and {Marchesini}, Danilo and {Wechsler}, Risa H. and {Muzzin}, Adam and {Papovich}, Casey and {Stefanon}, Mauro},
        title = "{Using Cumulative Number Densities to Compare Galaxies across Cosmic Time}",
      journal = {\apjl},
         year = 2013,
        month = nov,
       volume = {777},
       number = {1},
          eid = {L10},
        pages = {L10},
          doi = {10.1088/2041-8205/777/1/L10},
archivePrefix = {arXiv},
       eprint = {1308.3232},
 primaryClass = {astro-ph.CO},
       adsurl = {https://ui.adsabs.harvard.edu/abs/2013ApJ...777L..10B}
}

@ARTICLE{2014ApJ...793...12B,
       author = {{Birrer}, Simon and {Lilly}, Simon and {Amara}, Adam and {Paranjape}, Aseem and {Refregier}, Alexandre},
        title = "{A Simple Model Linking Galaxy and Dark Matter Evolution}",
      journal = {\apj},
         year = 2014,
        month = sep,
       volume = {793},
       number = {1},
          eid = {12},
        pages = {12},
          doi = {10.1088/0004-637X/793/1/12},
archivePrefix = {arXiv},
       eprint = {1401.3162},
 primaryClass = {astro-ph.CO},
       adsurl = {https://ui.adsabs.harvard.edu/abs/2014ApJ...793...12B}
}

@ARTICLE{2019MNRAS.488.3143B,
       author = {{Behroozi}, Peter and {Wechsler}, Risa H. and {Hearin}, Andrew P. and {Conroy}, Charlie},
        title = "{UNIVERSEMACHINE: The correlation between galaxy growth and dark matter halo assembly from z = 0-10}",
      journal = {\mnras},
         year = 2019,
        month = sep,
       volume = {488},
       number = {3},
        pages = {3143-3194},
          doi = {10.1093/mnras/stz1182},
archivePrefix = {arXiv},
       eprint = {1806.07893},
 primaryClass = {astro-ph.GA},
       adsurl = {https://ui.adsabs.harvard.edu/abs/2019MNRAS.488.3143B}
}

@ARTICLE{2024ApJ...971...69D,
       author = {{de Is{\'\i}dio}, Natanael G. and {Men{\'e}ndez-Delmestre}, K. and {Gon{\c{c}}alves}, T.~S. and {Grossi}, M. and {Rodrigues}, D.~C. and {Garavito-Camargo}, N. and {Araujo-Carvalho}, A. and {Beaklini}, P.~P.~B. and {Cavalcante-Coelho}, Y. and {Cortesi}, A. and {Quiroga-Nu{\~n}ez}, L.~H. and {Randriamampandry}, T.},
        title = "{Dark Matter Distribution in Milky Way analog Galaxies}",
      journal = {\apj},
         year = 2024,
        month = aug,
       volume = {971},
       number = {1},
          eid = {69},
        pages = {69},
          doi = {10.3847/1538-4357/ad53c8},
archivePrefix = {arXiv},
       eprint = {2310.13839},
 primaryClass = {astro-ph.GA},
       adsurl = {https://ui.adsabs.harvard.edu/abs/2024ApJ...971...69D}
}

@ARTICLE{2025A&A...699A.311M,
       author = {{Mancera Pi{\~n}a}, Pavel E. and {Read}, Justin I. and {Kim}, Stacy and {Marasco}, Antonino and {Benavides}, Jos{\'e} A. and {Glowacki}, Marcin and {Pezzulli}, Gabriele and {Lagos}, Claudia del P.},
        title = "{The galaxy-halo connection of disc galaxies over six orders of magnitude in stellar mass}",
      journal = {\aap},
         year = 2025,
        month = jul,
       volume = {699},
          eid = {A311},
        pages = {A311},
          doi = {10.1051/0004-6361/202554381},
archivePrefix = {arXiv},
       eprint = {2505.22727},
 primaryClass = {astro-ph.GA},
       adsurl = {https://ui.adsabs.harvard.edu/abs/2025A&A...699A.311M}
}

@ARTICLE{2018AstL...44....8K,
       author = {{Kravtsov}, A.~V. and {Vikhlinin}, A.~A. and {Meshcheryakov}, A.~V.},
        title = "{Stellar Mass{\textemdash}Halo Mass Relation and Star Formation Efficiency in High-Mass Halos}",
      journal = {Astronomy Letters},
         year = 2018,
        month = jan,
       volume = {44},
       number = {1},
        pages = {8-34},
          doi = {10.1134/S1063773717120015},
archivePrefix = {arXiv},
       eprint = {1401.7329},
 primaryClass = {astro-ph.CO},
       adsurl = {https://ui.adsabs.harvard.edu/abs/2018AstL...44....8K}
}

@ARTICLE{2019A&A...631A.175E,
       author = {{Erfanianfar}, G. and {Finoguenov}, A. and {Furnell}, K. and {Popesso}, P. and {Biviano}, A. and {Wuyts}, S. and {Collins}, C.~A. and {Mirkazemi}, M. and {Comparat}, J. and {Khosroshahi}, H. and {Nandra}, K. and {Capasso}, R. and {Rykoff}, E. and {Wilman}, D. and {Merloni}, A. and {Clerc}, N. and {Salvato}, M. and {Chitham}, J.~I. and {Kelvin}, L.~S. and {Gozaliasl}, G. and {Weijmans}, A. and {Brownstein}, J. and {Egami}, E. and {Pereira}, M.~J. and {Schneider}, D.~P. and {Kirkpatrick}, C. and {Damsted}, S. and {Kukkola}, A.},
        title = "{Stellar mass-halo mass relation for the brightest central galaxies of X-ray clusters since z {\ensuremath{\sim}} 0.65}",
      journal = {\aap},
         year = 2019,
        month = nov,
       volume = {631},
          eid = {A175},
        pages = {A175},
          doi = {10.1051/0004-6361/201935375},
archivePrefix = {arXiv},
       eprint = {1908.01559},
 primaryClass = {astro-ph.GA},
       adsurl = {https://ui.adsabs.harvard.edu/abs/2019A&A...631A.175E}
}

@ARTICLE{2022ApJ...928...28G,
       author = {{Golden-Marx}, Jesse B. and {Miller}, C.~J. and {Zhang}, Y. and {Ogando}, R.~L.~C. and {Palmese}, A. and {Abbott}, T.~M.~C. and {Aguena}, M. and {Allam}, S. and {Andrade-Oliveira}, F. and {Annis}, J. and {Bacon}, D. and {Bertin}, E. and {Brooks}, D. and {Buckley-Geer}, E. and {Carnero Rosell}, A. and {Carrasco Kind}, M. and {Castander}, F.~J. and {Costanzi}, M. and {Crocce}, M. and {da Costa}, L.~N. and {Pereira}, M.~E.~S. and {De Vicente}, J. and {Desai}, S. and {Diehl}, H.~T. and {Doel}, P. and {Drlica-Wagner}, A. and {Everett}, S. and {Evrard}, A.~E. and {Ferrero}, I. and {Flaugher}, B. and {Fosalba}, P. and {Frieman}, J. and {Garc{\'\i}a-Bellido}, J. and {Gaztanaga}, E. and {Gerdes}, D.~W. and {Gruen}, D. and {Gruendl}, R.~A. and {Gschwend}, J. and {Gutierrez}, G. and {Hartley}, W.~G. and {Hinton}, S.~R. and {Hollowood}, D.~L. and {Honscheid}, K. and {Hoyle}, B. and {James}, D.~J. and {Jeltema}, T. and {Kim}, A.~G. and {Krause}, E. and {Kuehn}, K. and {Kuropatkin}, N. and {Lahav}, O. and {Lima}, M. and {Maia}, M.~A.~G. and {Marshall}, J.~L. and {Melchior}, P. and {Menanteau}, F. and {Miquel}, R. and {Mohr}, J.~J. and {Morgan}, R. and {Paz-Chinch{\'o}n}, F. and {Petravick}, D. and {Pieres}, A. and {Plazas Malag{\'o}n}, A.~A. and {Prat}, J. and {Romer}, A.~K. and {Sanchez}, E. and {Santiago}, B. and {Scarpine}, V. and {Schubnell}, M. and {Serrano}, S. and {Sevilla-Noarbe}, I. and {Smith}, M. and {Soares-Santos}, M. and {Suchyta}, E. and {Tarle}, G. and {Varga}, T.~N.},
        title = "{The Observed Evolution of the Stellar Mass-Halo Mass Relation for Brightest Central Galaxies}",
      journal = {\apj},
         year = 2022,
        month = mar,
       volume = {928},
       number = {1},
          eid = {28},
        pages = {28},
          doi = {10.3847/1538-4357/ac4cb4},
archivePrefix = {arXiv},
       eprint = {2107.02197},
 primaryClass = {astro-ph.GA},
       adsurl = {https://ui.adsabs.harvard.edu/abs/2022ApJ...928...28G}
}

@ARTICLE{2014A&A...561A..79V,
       author = {{van der Burg}, R.~F.~J. and {Muzzin}, A. and {Hoekstra}, H. and {Wilson}, G. and {Lidman}, C. and {Yee}, H.~K.~C.},
        title = "{A census of stellar mass in ten massive haloes at z \raisebox{-0.5ex}\textasciitilde 1 from the GCLASS Survey}",
      journal = {\aap},
         year = 2014,
        month = jan,
       volume = {561},
          eid = {A79},
        pages = {A79},
          doi = {10.1051/0004-6361/201322771},
archivePrefix = {arXiv},
       eprint = {1310.0020},
 primaryClass = {astro-ph.CO},
       adsurl = {https://ui.adsabs.harvard.edu/abs/2014A&A...561A..79V}
}

@ARTICLE{2011MNRAS.410..210M,
       author = {{More}, Surhud and {van den Bosch}, Frank C. and {Cacciato}, Marcello and {Skibba}, Ramin and {Mo}, H.~J. and {Yang}, Xiaohu},
        title = "{Satellite kinematics - III. Halo masses of central galaxies in SDSS}",
      journal = {\mnras},
         year = 2011,
        month = jan,
       volume = {410},
       number = {1},
        pages = {210-226},
          doi = {10.1111/j.1365-2966.2010.17436.x},
archivePrefix = {arXiv},
       eprint = {1003.3203},
 primaryClass = {astro-ph.CO},
       adsurl = {https://ui.adsabs.harvard.edu/abs/2011MNRAS.410..210M}
}

@ARTICLE{2016MNRAS.459.3251V,
       author = {{van Uitert}, Edo and {Cacciato}, Marcello and {Hoekstra}, Henk and {Brouwer}, Margot and {Sif{\'o}n}, Crist{\'o}bal and {Viola}, Massimo and {Baldry}, Ivan and {Bland-Hawthorn}, Joss and {Brough}, Sarah and {Brown}, M.~J.~I. and {Choi}, Ami and {Driver}, Simon P. and {Erben}, Thomas and {Heymans}, Catherine and {Hildebrandt}, Hendrik and {Joachimi}, Benjamin and {Kuijken}, Konrad and {Liske}, Jochen and {Loveday}, Jon and {McFarland}, John and {Miller}, Lance and {Nakajima}, Reiko and {Peacock}, John and {Radovich}, Mario and {Robotham}, A.~S.~G. and {Schneider}, Peter and {Sikkema}, Gert and {Taylor}, Edward N. and {Verdoes Kleijn}, Gijs},
        title = "{The stellar-to-halo mass relation of GAMA galaxies from 100 deg$^{2}$ of KiDS weak lensing data}",
      journal = {\mnras},
         year = 2016,
        month = jul,
       volume = {459},
       number = {3},
        pages = {3251-3270},
          doi = {10.1093/mnras/stw747},
archivePrefix = {arXiv},
       eprint = {1601.06791},
 primaryClass = {astro-ph.GA},
       adsurl = {https://ui.adsabs.harvard.edu/abs/2016MNRAS.459.3251V}
}

@ARTICLE{2013ApJ...771...30R,
       author = {{Reddick}, Rachel M. and {Wechsler}, Risa H. and {Tinker}, Jeremy L. and {Behroozi}, Peter S.},
        title = "{The Connection between Galaxies and Dark Matter Structures in the Local Universe}",
      journal = {\apj},
         year = 2013,
        month = jul,
       volume = {771},
       number = {1},
          eid = {30},
        pages = {30},
          doi = {10.1088/0004-637X/771/1/30},
archivePrefix = {arXiv},
       eprint = {1207.2160},
 primaryClass = {astro-ph.CO},
       adsurl = {https://ui.adsabs.harvard.edu/abs/2013ApJ...771...30R}
}

@ARTICLE{2017ApJ...840...34S,
       author = {{Shankar}, Francesco and {Sonnenfeld}, Alessandro and {Mamon}, Gary A. and {Chae}, Kyu-Hyun and {Gavazzi}, Raphael and {Treu}, Tommaso and {Diemer}, Benedikt and {Nipoti}, Carlo and {Buchan}, Stewart and {Bernardi}, Mariangela and {Sheth}, Ravi and {Huertas-Company}, Marc},
        title = "{Revisiting the Bulge-Halo Conspiracy. I. Dependence on Galaxy Properties and Halo Mass}",
      journal = {\apj},
         year = 2017,
        month = may,
       volume = {840},
       number = {1},
          eid = {34},
        pages = {34},
          doi = {10.3847/1538-4357/aa66ce},
archivePrefix = {arXiv},
       eprint = {1703.06145},
 primaryClass = {astro-ph.GA},
       adsurl = {https://ui.adsabs.harvard.edu/abs/2017ApJ...840...34S}
}

@ARTICLE{2020A&A...642A..83D,
       author = {{Dvornik}, Andrej and {Hoekstra}, Henk and {Kuijken}, Konrad and {Wright}, Angus H. and {Asgari}, Marika and {Bilicki}, Maciej and {Erben}, Thomas and {Giblin}, Benjamin and {Graham}, Alister W. and {Heymans}, Catherine and {Hildebrandt}, Hendrik and {Hopkins}, Andrew M. and {Kannawadi}, Arun and {Lin}, Chieh-An and {Taylor}, Edward N. and {Tr{\"o}ster}, Tilman},
        title = "{KiDS+GAMA: The weak lensing calibrated stellar-to-halo mass relation of central and satellite galaxies}",
      journal = {\aap},
         year = 2020,
        month = oct,
       volume = {642},
          eid = {A83},
        pages = {A83},
          doi = {10.1051/0004-6361/202038693},
archivePrefix = {arXiv},
       eprint = {2006.10777},
 primaryClass = {astro-ph.CO},
       adsurl = {https://ui.adsabs.harvard.edu/abs/2020A&A...642A..83D}
}

@ARTICLE{2015MNRAS.447..298H,
       author = {{Hudson}, Michael J. and {Gillis}, Bryan R. and {Coupon}, Jean and {Hildebrandt}, Hendrik and {Erben}, Thomas and {Heymans}, Catherine and {Hoekstra}, Henk and {Kitching}, Thomas D. and {Mellier}, Yannick and {Miller}, Lance and {Van Waerbeke}, Ludovic and {Bonnett}, Christopher and {Fu}, Liping and {Kuijken}, Konrad and {Rowe}, Barnaby and {Schrabback}, Tim and {Semboloni}, Elisabetta and {van Uitert}, Edo and {Velander}, Malin},
        title = "{CFHTLenS: co-evolution of galaxies and their dark matter haloes}",
      journal = {\mnras},
         year = 2015,
        month = feb,
       volume = {447},
       number = {1},
        pages = {298-314},
          doi = {10.1093/mnras/stu2367},
archivePrefix = {arXiv},
       eprint = {1310.6784},
 primaryClass = {astro-ph.CO},
       adsurl = {https://ui.adsabs.harvard.edu/abs/2015MNRAS.447..298H}
}

@ARTICLE{2006MNRAS.368..715M,
       author = {{Mandelbaum}, Rachel and {Seljak}, Uro{\v{s}} and {Kauffmann}, Guinevere and {Hirata}, Christopher M. and {Brinkmann}, Jonathan},
        title = "{Galaxy halo masses and satellite fractions from galaxy-galaxy lensing in the Sloan Digital Sky Survey: stellar mass, luminosity, morphology and environment dependencies}",
      journal = {\mnras},
         year = 2006,
        month = may,
       volume = {368},
       number = {2},
        pages = {715-731},
          doi = {10.1111/j.1365-2966.2006.10156.x},
archivePrefix = {arXiv},
       eprint = {astro-ph/0511164},
 primaryClass = {astro-ph},
       adsurl = {https://ui.adsabs.harvard.edu/abs/2006MNRAS.368..715M}
}

@ARTICLE{2011A&A...534A..14V,
       author = {{van Uitert}, E. and {Hoekstra}, H. and {Velander}, M. and {Gilbank}, D.~G. and {Gladders}, M.~D. and {Yee}, H.~K.~C.},
        title = "{Galaxy-galaxy lensing constraints on the relation between baryons and dark matter in galaxies in the Red Sequence Cluster Survey 2}",
      journal = {\aap},
         year = 2011,
        month = oct,
       volume = {534},
          eid = {A14},
        pages = {A14},
          doi = {10.1051/0004-6361/201117308},
archivePrefix = {arXiv},
       eprint = {1107.4093},
 primaryClass = {astro-ph.CO},
       adsurl = {https://ui.adsabs.harvard.edu/abs/2011A&A...534A..14V}
}

@ARTICLE{2012MNRAS.425.2610R,
       author = {{Reyes}, R. and {Mandelbaum}, R. and {Gunn}, J.~E. and {Nakajima}, R. and {Seljak}, U. and {Hirata}, C.~M.},
        title = "{Optical-to-virial velocity ratios of local disc galaxies from combined kinematics and galaxy-galaxy lensing}",
      journal = {\mnras},
         year = 2012,
        month = oct,
       volume = {425},
       number = {4},
        pages = {2610-2640},
          doi = {10.1111/j.1365-2966.2012.21472.x},
archivePrefix = {arXiv},
       eprint = {1110.4107},
 primaryClass = {astro-ph.CO},
       adsurl = {https://ui.adsabs.harvard.edu/abs/2012MNRAS.425.2610R}
}

@ARTICLE{2025A&A...704A.110C,
       author = {{Chiu}, I.-Non and {Ghirardini}, Vittorio and {Grandis}, Sebastian and {Okabe}, Nobuhiro and {Artis}, Emmanuel and {Bulbul}, Esra and {Emre Bahar}, Y. and {Balzer}, Fabian and {Clerc}, Nicolas and {Comparat}, Johan and {Hsieh}, Bau-Ching and {Kleinebreil}, Florian and {Kluge}, Matthias and {Liu}, Ang and {Monteiro-Oliveira}, Rog{\'e}rio and {Oguri}, Masamune and {Pacaud}, Florian and {Ramos Ceja}, Miriam and {Reiprich}, Thomas H. and {Sanders}, Jeremy and {Schrabback}, Tim and {Seppi}, Riccardo and {Sommer}, Martin and {Tam}, Sut-Ieng and {Umetsu}, Keiichi and {Zhang}, Xiaoyuan},
        title = "{The SRG/eROSITA All-Sky Survey: The weak-lensing mass calibration and the stellar mass-to-halo mass relation from the Hyper Suprime-Cam Subaru Strategic Program}",
      journal = {\aap},
         year = 2025,
        month = dec,
       volume = {704},
          eid = {A110},
        pages = {A110},
          doi = {10.1051/0004-6361/202554942},
archivePrefix = {arXiv},
       eprint = {2504.01076},
 primaryClass = {astro-ph.CO},
       adsurl = {https://ui.adsabs.harvard.edu/abs/2025A&A...704A.110C}
}

@ARTICLE{2022PASJ...74..175A,
       author = {{Akino}, Daichi and {Eckert}, Dominique and {Okabe}, Nobuhiro and {Sereno}, Mauro and {Umetsu}, Keiichi and {Oguri}, Masamune and {Gastaldello}, Fabio and {Chiu}, I. -Non and {Ettori}, Stefano and {Evrard}, August E. and {Farahi}, Arya and {Maughan}, Ben and {Pierre}, Marguerite and {Ricci}, Marina and {Valtchanov}, Ivan and {McCarthy}, Ian and {McGee}, Sean and {Miyazaki}, Satoshi and {Nishizawa}, Atsushi J. and {Tanaka}, Masayuki},
        title = "{HSC-XXL: Baryon budget of the 136 XXL groups and clusters}",
      journal = {\pasj},
         year = 2022,
        month = feb,
       volume = {74},
       number = {1},
        pages = {175-208},
          doi = {10.1093/pasj/psab115},
archivePrefix = {arXiv},
       eprint = {2111.10080},
 primaryClass = {astro-ph.CO},
       adsurl = {https://ui.adsabs.harvard.edu/abs/2022PASJ...74..175A}
}

@ARTICLE{2013ApJ...778...14G,
       author = {{Gonzalez}, Anthony H. and {Sivanandam}, Suresh and {Zabludoff}, Ann I. and {Zaritsky}, Dennis},
        title = "{Galaxy Cluster Baryon Fractions Revisited}",
      journal = {\apj},
         year = 2013,
        month = nov,
       volume = {778},
       number = {1},
          eid = {14},
        pages = {14},
          doi = {10.1088/0004-637X/778/1/14},
archivePrefix = {arXiv},
       eprint = {1309.3565},
 primaryClass = {astro-ph.CO},
       adsurl = {https://ui.adsabs.harvard.edu/abs/2013ApJ...778...14G}
}

@ARTICLE{1994MNRAS.268..405N,
       author = {{Nandra}, K. and {Pounds}, K.~A.},
        title = "{GINGA observations of the X-ray spectra of Seyfert galaxies.}",
      journal = {\mnras},
         year = 1994,
        month = may,
       volume = {268},
        pages = {405-429},
          doi = {10.1093/mnras/268.2.405},
       adsurl = {https://ui.adsabs.harvard.edu/abs/1994MNRAS.268..405N}
}

@ARTICLE{2010MNRAS.404.1111G,
       author = {{Guo}, Qi and {White}, Simon and {Li}, Cheng and {Boylan-Kolchin}, Michael},
        title = "{How do galaxies populate dark matter haloes?}",
      journal = {\mnras},
         year = 2010,
        month = may,
       volume = {404},
       number = {3},
        pages = {1111-1120},
          doi = {10.1111/j.1365-2966.2010.16341.x},
archivePrefix = {arXiv},
       eprint = {0909.4305},
 primaryClass = {astro-ph.CO},
       adsurl = {https://ui.adsabs.harvard.edu/abs/2010MNRAS.404.1111G}
}

@ARTICLE{1998A&A...331L...1S,
       author = {{Silk}, Joseph and {Rees}, Martin J.},
        title = "{Quasars and galaxy formation}",
      journal = {\aap},
         year = 1998,
        month = mar,
       volume = {331},
        pages = {L1-L4},
          doi = {10.48550/arXiv.astro-ph/9801013},
archivePrefix = {arXiv},
       eprint = {astro-ph/9801013},
 primaryClass = {astro-ph},
       adsurl = {https://ui.adsabs.harvard.edu/abs/1998A&A...331L...1S}
}

@ARTICLE{2007ARA&A..45..117M,
       author = {{McNamara}, B.~R. and {Nulsen}, P.~E.~J.},
        title = "{Heating Hot Atmospheres with Active Galactic Nuclei}",
      journal = {\araa},
         year = 2007,
        month = sep,
       volume = {45},
       number = {1},
        pages = {117-175},
          doi = {10.1146/annurev.astro.45.051806.110625},
archivePrefix = {arXiv},
       eprint = {0709.2152},
 primaryClass = {astro-ph},
       adsurl = {https://ui.adsabs.harvard.edu/abs/2007ARA&A..45..117M}
}

@ARTICLE{2012ARA&A..50..353K,
       author = {{Kravtsov}, Andrey V. and {Borgani}, Stefano},
        title = "{Formation of Galaxy Clusters}",
      journal = {\araa},
         year = 2012,
        month = sep,
       volume = {50},
        pages = {353-409},
          doi = {10.1146/annurev-astro-081811-125502},
archivePrefix = {arXiv},
       eprint = {1205.5556},
 primaryClass = {astro-ph.CO},
       adsurl = {https://ui.adsabs.harvard.edu/abs/2012ARA&A..50..353K}
}

@ARTICLE{2016MNRAS.457.3200M,
       author = {{Mandelbaum}, Rachel and {Wang}, Wenting and {Zu}, Ying and {White}, Simon and {Henriques}, Bruno and {More}, Surhud},
        title = "{Strong bimodality in the host halo mass of central galaxies from galaxy-galaxy lensing}",
      journal = {\mnras},
         year = 2016,
        month = apr,
       volume = {457},
       number = {3},
        pages = {3200-3218},
          doi = {10.1093/mnras/stw188},
archivePrefix = {arXiv},
       eprint = {1509.06762},
 primaryClass = {astro-ph.GA},
       adsurl = {https://ui.adsabs.harvard.edu/abs/2016MNRAS.457.3200M}
}

@ARTICLE{2005ApJ...635...73H,
       author = {{Hoekstra}, H. and {Hsieh}, B.~C. and {Yee}, H.~K.~C. and {Lin}, H. and {Gladders}, M.~D.},
        title = "{Virial Masses and the Baryon Fraction in Galaxies}",
      journal = {\apj},
         year = 2005,
        month = dec,
       volume = {635},
       number = {1},
        pages = {73-85},
          doi = {10.1086/496913},
archivePrefix = {arXiv},
       eprint = {astro-ph/0510097},
 primaryClass = {astro-ph},
       adsurl = {https://ui.adsabs.harvard.edu/abs/2005ApJ...635...73H}
}

@ARTICLE{2016MNRAS.455..258C,
       author = {{Chiu}, I. and {Mohr}, J. and {McDonald}, M. and {Bocquet}, S. and {Ashby}, M.~L.~N. and {Bayliss}, M. and {Benson}, B.~A. and {Bleem}, L.~E. and {Brodwin}, M. and {Desai}, S. and {Dietrich}, J.~P. and {Forman}, W.~R. and {Gangkofner}, C. and {Gonzalez}, A.~H. and {Hennig}, C. and {Liu}, J. and {Reichardt}, C.~L. and {Saro}, A. and {Stalder}, B. and {Stanford}, S.~A. and {Song}, J. and {Schrabback}, T. and {{\v{S}}uhada}, R. and {Strazzullo}, V. and {Zenteno}, A.},
        title = "{Baryon content of massive galaxy clusters at 0.57 < z < 1.33}",
      journal = {\mnras},
         year = 2016,
        month = jan,
       volume = {455},
       number = {1},
        pages = {258-275},
          doi = {10.1093/mnras/stv2303},
archivePrefix = {arXiv},
       eprint = {1412.7823},
 primaryClass = {astro-ph.CO},
       adsurl = {https://ui.adsabs.harvard.edu/abs/2016MNRAS.455..258C}
}

@ARTICLE{2004ApJ...617..879L,
       author = {{Lin}, Yen-Ting and {Mohr}, Joseph J.},
        title = "{K-band Properties of Galaxy Clusters and Groups: Brightest Cluster Galaxies and Intracluster Light}",
      journal = {\apj},
         year = 2004,
        month = dec,
       volume = {617},
       number = {2},
        pages = {879-895},
          doi = {10.1086/425412},
archivePrefix = {arXiv},
       eprint = {astro-ph/0408557},
 primaryClass = {astro-ph},
       adsurl = {https://ui.adsabs.harvard.edu/abs/2004ApJ...617..879L}
}

@ARTICLE{2018ApJ...860....2G,
       author = {{Golden-Marx}, Jesse B. and {Miller}, Christopher J.},
        title = "{The Impact of Environment on the Stellar Mass-Halo Mass Relation}",
      journal = {\apj},
         year = 2018,
        month = jun,
       volume = {860},
       number = {1},
          eid = {2},
        pages = {2},
          doi = {10.3847/1538-4357/aac2bd},
archivePrefix = {arXiv},
       eprint = {1711.00481},
 primaryClass = {astro-ph.GA},
       adsurl = {https://ui.adsabs.harvard.edu/abs/2018ApJ...860....2G}
}

@ARTICLE{2009ApJ...699.1333H,
       author = {{Hansen}, Sarah M. and {Sheldon}, Erin S. and {Wechsler}, Risa H. and {Koester}, Benjamin P.},
        title = "{The Galaxy Content of SDSS Clusters and Groups}",
      journal = {\apj},
         year = 2009,
        month = jul,
       volume = {699},
       number = {2},
        pages = {1333-1353},
          doi = {10.1088/0004-637X/699/2/1333},
archivePrefix = {arXiv},
       eprint = {0710.3780},
 primaryClass = {astro-ph},
       adsurl = {https://ui.adsabs.harvard.edu/abs/2009ApJ...699.1333H}
}

@ARTICLE{2019ApJ...878...14G,
       author = {{Golden-Marx}, Jesse B. and {Miller}, Christopher J.},
        title = "{The Impact of Environment on Late-time Evolution of the Stellar Mass-Halo Mass Relation}",
      journal = {\apj},
         year = 2019,
        month = jun,
       volume = {878},
       number = {1},
          eid = {14},
        pages = {14},
          doi = {10.3847/1538-4357/ab1d55},
archivePrefix = {arXiv},
       eprint = {1901.02568},
 primaryClass = {astro-ph.GA},
       adsurl = {https://ui.adsabs.harvard.edu/abs/2019ApJ...878...14G}
}

@ARTICLE{2012ApJ...752...41Y,
       author = {{Yang}, Xiaohu and {Mo}, H.~J. and {van den Bosch}, Frank C. and {Zhang}, Youcai and {Han}, Jiaxin},
        title = "{Evolution of the Galaxy-Dark Matter Connection and the Assembly of Galaxies in Dark Matter Halos}",
      journal = {\apj},
         year = 2012,
        month = jun,
       volume = {752},
       number = {1},
          eid = {41},
        pages = {41},
          doi = {10.1088/0004-637X/752/1/41},
archivePrefix = {arXiv},
       eprint = {1110.1420},
 primaryClass = {astro-ph.CO},
       adsurl = {https://ui.adsabs.harvard.edu/abs/2012ApJ...752...41Y}
}

@ARTICLE{2018MNRAS.474.3009D,
       author = {{DeMaio}, Tahlia and {Gonzalez}, Anthony H. and {Zabludoff}, Ann and {Zaritsky}, Dennis and {Connor}, Thomas and {Donahue}, Megan and {Mulchaey}, John S.},
        title = "{Lost but not forgotten: intracluster light in galaxy groups and clusters}",
      journal = {\mnras},
         year = 2018,
        month = mar,
       volume = {474},
       number = {3},
        pages = {3009-3031},
          doi = {10.1093/mnras/stx2946},
archivePrefix = {arXiv},
       eprint = {1710.11313},
 primaryClass = {astro-ph.GA},
       adsurl = {https://ui.adsabs.harvard.edu/abs/2018MNRAS.474.3009D}
}

@ARTICLE{2024A&A...686A.157N,
       author = {{Nelson}, Dylan and {Pillepich}, Annalisa and {Ayromlou}, Mohammadreza and {Lee}, Wonki and {Lehle}, Katrin and {Rohr}, Eric and {Truong}, Nhut},
        title = "{Introducing the TNG-Cluster simulation: Overview and the physical properties of the gaseous intracluster medium}",
      journal = {\aap},
         year = 2024,
        month = jun,
       volume = {686},
          eid = {A157},
        pages = {A157},
          doi = {10.1051/0004-6361/202348608},
archivePrefix = {arXiv},
       eprint = {2311.06338},
 primaryClass = {astro-ph.GA},
       adsurl = {https://ui.adsabs.harvard.edu/abs/2024A&A...686A.157N}
}

@ARTICLE{2024A&A...686A..86R,
       author = {{Rohr}, Eric and {Pillepich}, Annalisa and {Nelson}, Dylan and {Ayromlou}, Mohammadreza and {Zinger}, Elad},
        title = "{The hot circumgalactic media of massive cluster satellites in the TNG-Cluster simulation: Existence and detectability}",
      journal = {\aap},
         year = 2024,
        month = jun,
       volume = {686},
          eid = {A86},
        pages = {A86},
          doi = {10.1051/0004-6361/202348583},
archivePrefix = {arXiv},
       eprint = {2311.06337},
 primaryClass = {astro-ph.GA},
       adsurl = {https://ui.adsabs.harvard.edu/abs/2024A&A...686A..86R}
}

@ARTICLE{2018MNRAS.477.1206N,
       author = {{Naiman}, Jill P. and {Pillepich}, Annalisa and {Springel}, Volker and {Ramirez-Ruiz}, Enrico and {Torrey}, Paul and {Vogelsberger}, Mark and {Pakmor}, R{\"u}diger and {Nelson}, Dylan and {Marinacci}, Federico and {Hernquist}, Lars and {Weinberger}, Rainer and {Genel}, Shy},
        title = "{First results from the IllustrisTNG simulations: a tale of two elements - chemical evolution of magnesium and europium}",
      journal = {\mnras},
         year = 2018,
        month = jun,
       volume = {477},
       number = {1},
        pages = {1206-1224},
          doi = {10.1093/mnras/sty618},
archivePrefix = {arXiv},
       eprint = {1707.03401},
 primaryClass = {astro-ph.GA},
       adsurl = {https://ui.adsabs.harvard.edu/abs/2018MNRAS.477.1206N}
}

@ARTICLE{2018MNRAS.475..648P,
       author = {{Pillepich}, Annalisa and {Nelson}, Dylan and {Hernquist}, Lars and {Springel}, Volker and {Pakmor}, R{\"u}diger and {Torrey}, Paul and {Weinberger}, Rainer and {Genel}, Shy and {Naiman}, Jill P. and {Marinacci}, Federico and {Vogelsberger}, Mark},
        title = "{First results from the IllustrisTNG simulations: the stellar mass content of groups and clusters of galaxies}",
      journal = {\mnras},
         year = 2018,
        month = mar,
       volume = {475},
       number = {1},
        pages = {648-675},
          doi = {10.1093/mnras/stx3112},
archivePrefix = {arXiv},
       eprint = {1707.03406},
 primaryClass = {astro-ph.GA},
       adsurl = {https://ui.adsabs.harvard.edu/abs/2018MNRAS.475..648P}
}

@ARTICLE{2018MNRAS.475..676S,
       author = {{Springel}, Volker and {Pakmor}, R{\"u}diger and {Pillepich}, Annalisa and {Weinberger}, Rainer and {Nelson}, Dylan and {Hernquist}, Lars and {Vogelsberger}, Mark and {Genel}, Shy and {Torrey}, Paul and {Marinacci}, Federico and {Naiman}, Jill},
        title = "{First results from the IllustrisTNG simulations: matter and galaxy clustering}",
      journal = {\mnras},
         year = 2018,
        month = mar,
       volume = {475},
       number = {1},
        pages = {676-698},
          doi = {10.1093/mnras/stx3304},
archivePrefix = {arXiv},
       eprint = {1707.03397},
 primaryClass = {astro-ph.GA},
       adsurl = {https://ui.adsabs.harvard.edu/abs/2018MNRAS.475..676S}
}

@ARTICLE{2018MNRAS.480.5113M,
       author = {{Marinacci}, Federico and {Vogelsberger}, Mark and {Pakmor}, R{\"u}diger and {Torrey}, Paul and {Springel}, Volker and {Hernquist}, Lars and {Nelson}, Dylan and {Weinberger}, Rainer and {Pillepich}, Annalisa and {Naiman}, Jill and {Genel}, Shy},
        title = "{First results from the IllustrisTNG simulations: radio haloes and magnetic fields}",
      journal = {\mnras},
         year = 2018,
        month = nov,
       volume = {480},
       number = {4},
        pages = {5113-5139},
          doi = {10.1093/mnras/sty2206},
archivePrefix = {arXiv},
       eprint = {1707.03396},
 primaryClass = {astro-ph.CO},
       adsurl = {https://ui.adsabs.harvard.edu/abs/2018MNRAS.480.5113M}
}

@ARTICLE{2018MNRAS.475..624N,
       author = {{Nelson}, Dylan and {Pillepich}, Annalisa and {Springel}, Volker and {Weinberger}, Rainer and {Hernquist}, Lars and {Pakmor}, R{\"u}diger and {Genel}, Shy and {Torrey}, Paul and {Vogelsberger}, Mark and {Kauffmann}, Guinevere and {Marinacci}, Federico and {Naiman}, Jill},
        title = "{First results from the IllustrisTNG simulations: the galaxy colour bimodality}",
      journal = {\mnras},
         year = 2018,
        month = mar,
       volume = {475},
       number = {1},
        pages = {624-647},
          doi = {10.1093/mnras/stx3040},
archivePrefix = {arXiv},
       eprint = {1707.03395},
 primaryClass = {astro-ph.GA},
       adsurl = {https://ui.adsabs.harvard.edu/abs/2018MNRAS.475..624N}
}

@ARTICLE{2014MNRAS.444.1518V,
       author = {{Vogelsberger}, Mark and {Genel}, Shy and {Springel}, Volker and {Torrey}, Paul and {Sijacki}, Debora and {Xu}, Dandan and {Snyder}, Greg and {Nelson}, Dylan and {Hernquist}, Lars},
        title = "{Introducing the Illustris Project: simulating the coevolution of dark and visible matter in the Universe}",
      journal = {\mnras},
         year = 2014,
        month = oct,
       volume = {444},
       number = {2},
        pages = {1518-1547},
          doi = {10.1093/mnras/stu1536},
archivePrefix = {arXiv},
       eprint = {1405.2921},
 primaryClass = {astro-ph.CO},
       adsurl = {https://ui.adsabs.harvard.edu/abs/2014MNRAS.444.1518V}
}

@ARTICLE{2014MNRAS.445..175G,
       author = {{Genel}, Shy and {Vogelsberger}, Mark and {Springel}, Volker and {Sijacki}, Debora and {Nelson}, Dylan and {Snyder}, Greg and {Rodriguez-Gomez}, Vicente and {Torrey}, Paul and {Hernquist}, Lars},
        title = "{Introducing the Illustris project: the evolution of galaxy populations across cosmic time}",
      journal = {\mnras},
         year = 2014,
        month = nov,
       volume = {445},
       number = {1},
        pages = {175-200},
          doi = {10.1093/mnras/stu1654},
archivePrefix = {arXiv},
       eprint = {1405.3749},
 primaryClass = {astro-ph.CO},
       adsurl = {https://ui.adsabs.harvard.edu/abs/2014MNRAS.445..175G}
}

@ARTICLE{2015MNRAS.452..575S,
       author = {{Sijacki}, Debora and {Vogelsberger}, Mark and {Genel}, Shy and {Springel}, Volker and {Torrey}, Paul and {Snyder}, Gregory F. and {Nelson}, Dylan and {Hernquist}, Lars},
        title = "{The Illustris simulation: the evolving population of black holes across cosmic time}",
      journal = {\mnras},
         year = 2015,
        month = sep,
       volume = {452},
       number = {1},
        pages = {575-596},
          doi = {10.1093/mnras/stv1340},
archivePrefix = {arXiv},
       eprint = {1408.6842},
 primaryClass = {astro-ph.GA},
       adsurl = {https://ui.adsabs.harvard.edu/abs/2015MNRAS.452..575S}
}

@ARTICLE{2015MNRAS.450.1937C,
       author = {{Crain}, Robert A. and {Schaye}, Joop and {Bower}, Richard G. and {Furlong}, Michelle and {Schaller}, Matthieu and {Theuns}, Tom and {Dalla Vecchia}, Claudio and {Frenk}, Carlos S. and {McCarthy}, Ian G. and {Helly}, John C. and {Jenkins}, Adrian and {Rosas-Guevara}, Yetli M. and {White}, Simon D.~M. and {Trayford}, James W.},
        title = "{The EAGLE simulations of galaxy formation: calibration of subgrid physics and model variations}",
      journal = {\mnras},
         year = 2015,
        month = jun,
       volume = {450},
       number = {2},
        pages = {1937-1961},
          doi = {10.1093/mnras/stv725},
archivePrefix = {arXiv},
       eprint = {1501.01311},
 primaryClass = {astro-ph.GA},
       adsurl = {https://ui.adsabs.harvard.edu/abs/2015MNRAS.450.1937C}
}

@ARTICLE{2015MNRAS.446..521S,
       author = {{Schaye}, Joop and {Crain}, Robert A. and {Bower}, Richard G. and {Furlong}, Michelle and {Schaller}, Matthieu and {Theuns}, Tom and {Dalla Vecchia}, Claudio and {Frenk}, Carlos S. and {McCarthy}, I.~G. and {Helly}, John C. and {Jenkins}, Adrian and {Rosas-Guevara}, Y.~M. and {White}, Simon D.~M. and {Baes}, Maarten and {Booth}, C.~M. and {Camps}, Peter and {Navarro}, Julio F. and {Qu}, Yan and {Rahmati}, Alireza and {Sawala}, Till and {Thomas}, Peter A. and {Trayford}, James},
        title = "{The EAGLE project: simulating the evolution and assembly of galaxies and their environments}",
      journal = {\mnras},
         year = 2015,
        month = jan,
       volume = {446},
       number = {1},
        pages = {521-554},
          doi = {10.1093/mnras/stu2058},
archivePrefix = {arXiv},
       eprint = {1407.7040},
 primaryClass = {astro-ph.GA},
       adsurl = {https://ui.adsabs.harvard.edu/abs/2015MNRAS.446..521S}
}

@ARTICLE{2013MNRAS.428.2407W,
       author = {{Wojtak}, Rados{\l}aw and {Mamon}, Gary A.},
        title = "{Physical properties underlying observed kinematics of satellite galaxies}",
      journal = {\mnras},
         year = 2013,
        month = jan,
       volume = {428},
       number = {3},
        pages = {2407-2417},
          doi = {10.1093/mnras/sts203},
archivePrefix = {arXiv},
       eprint = {1207.1647},
 primaryClass = {astro-ph.CO},
       adsurl = {https://ui.adsabs.harvard.edu/abs/2013MNRAS.428.2407W}
}

@ARTICLE{Salim,
       author = {{Salim}, Samir and {Lee}, Janice C. and {Janowiecki}, Steven and {da Cunha}, Elisabete and {Dickinson}, Mark and {Boquien}, M{\'e}d{\'e}ric and {Burgarella}, Denis and {Salzer}, John J. and {Charlot}, St{\'e}phane},
        title = "{GALEX-SDSS-WISE Legacy Catalog (GSWLC): Star Formation Rates, Stellar Masses, and Dust Attenuations of 700,000 Low-redshift Galaxies}",
      journal = {\apjs},
         year = 2016,
        month = nov,
       volume = {227},
       number = {1},
          eid = {2},
        pages = {2},
          doi = {10.3847/0067-0049/227/1/2},
archivePrefix = {arXiv},
       eprint = {1610.00712},
 primaryClass = {astro-ph.GA},
       adsurl = {https://ui.adsabs.harvard.edu/abs/2016ApJS..227....2S}
}

\begin{appendix}

\section{Stacked X-ray spectra and spectral modeling results}\label{ap:spectra}

\begin{figure}[h!]
\centering
\includegraphics[width=0.93\hsize]{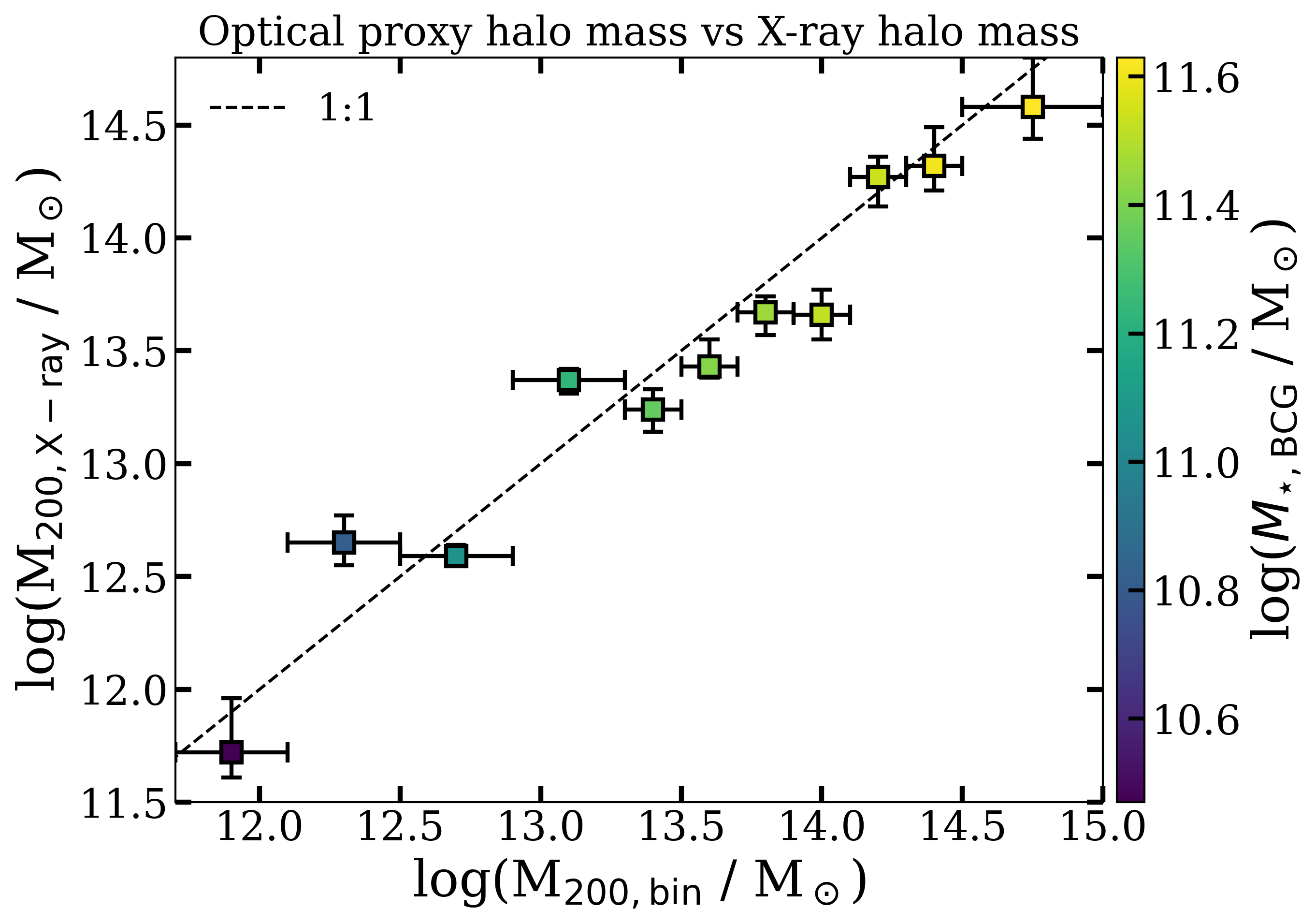}
  \caption{{X-ray-derived halo mass as a function of the average optically-based halo mass from \cite{yang_galaxy_2007} for bins of optical halo mass. The color of the points indicates the average stellar mass of the BCG.}
  \label{apf:mh_mh}}
\end{figure}

\begin{figure}[h!]
\centering
\includegraphics[width=0.93\hsize]{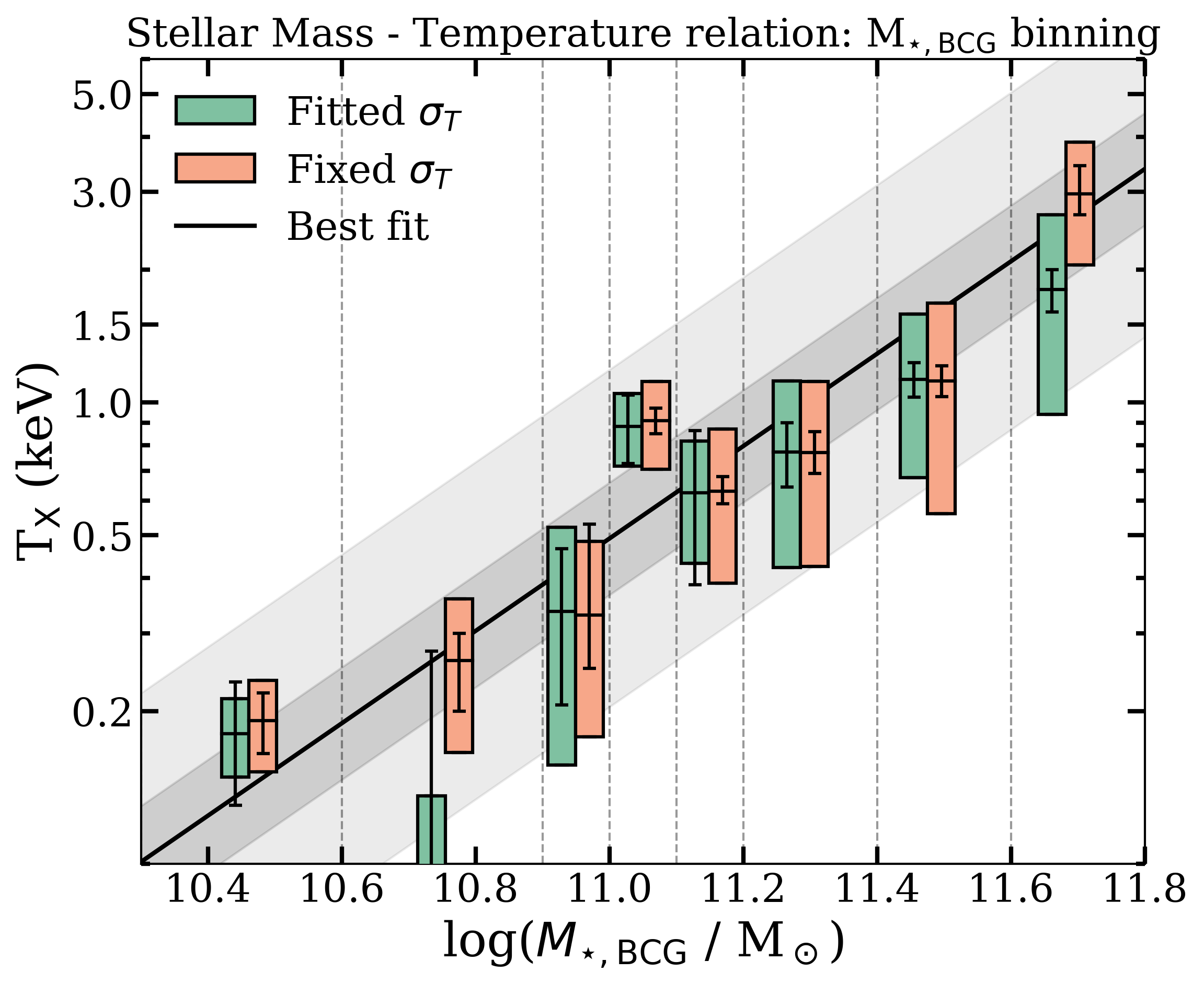}
  \caption{{Stellar mass-temperature relation for BCGs. Colored boxes show the measured X-ray temperature {$T_\mathrm{X}$} in each stellar-mass bin; the box center marks the {$T_\mathrm{X}$} value, while the box height represents the width of the temperature distribution $\sigma_T$ in the {\sc gadem} model. Orange boxes indicate predictions for $\sigma_T$ and the corresponding {$T_\mathrm{X}$} obtained from fits with $\sigma_T$ fixed to these values. Green boxes show results from fits with $\sigma_T$ left free. The solid line and shaded regions correspond to the best-fit relation from Eq.~\ref{eq:mstar_t} and its $1\sigma$ and $3\sigma$ uncertainties. Error bars indicate the uncertainties on {$T_\mathrm{X}$} for each fitting choice. Vertical lines mark the stellar-mass bin edges; within each bin, all boxes share the same $M_\star$, placed between the boxes.}
  \label{apf:tsigma}}
\end{figure}

\begin{figure*}[h!]
\centering
\includegraphics[width=0.33\hsize]{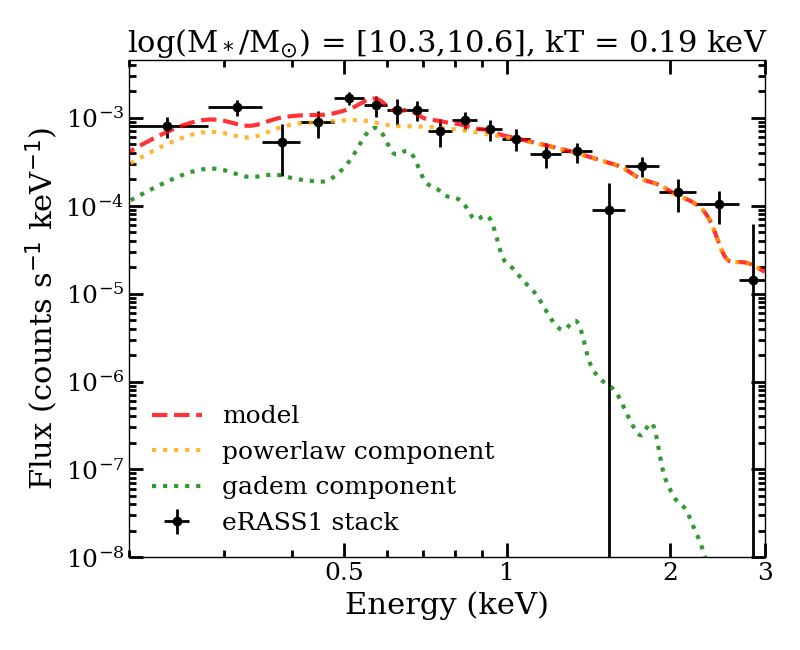}
\includegraphics[width=0.33\hsize]{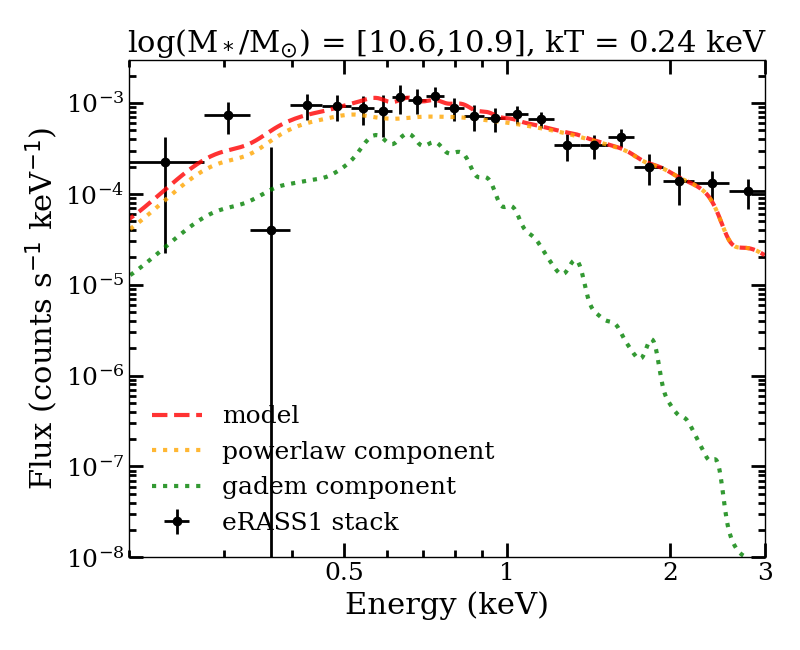}
\includegraphics[width=0.33\hsize]{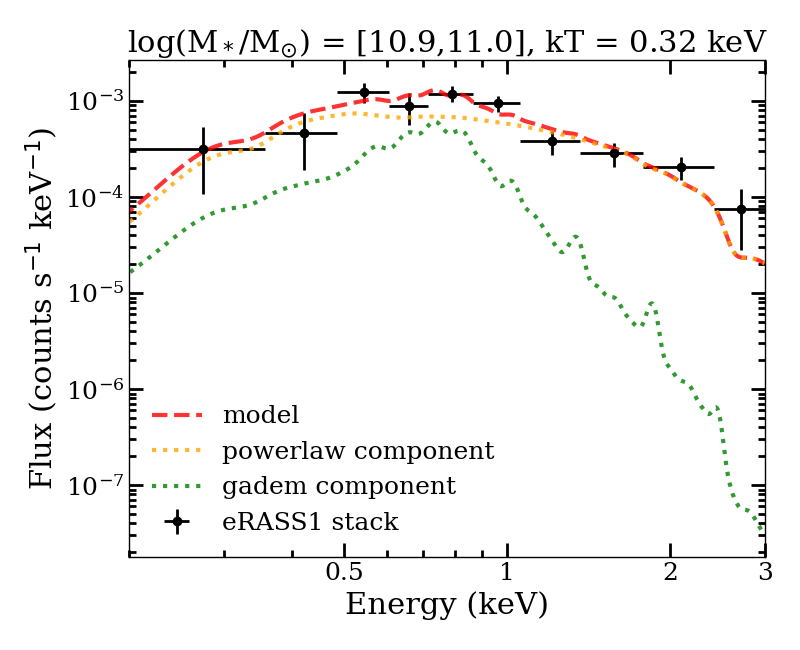}\\
\includegraphics[width=0.33\hsize]{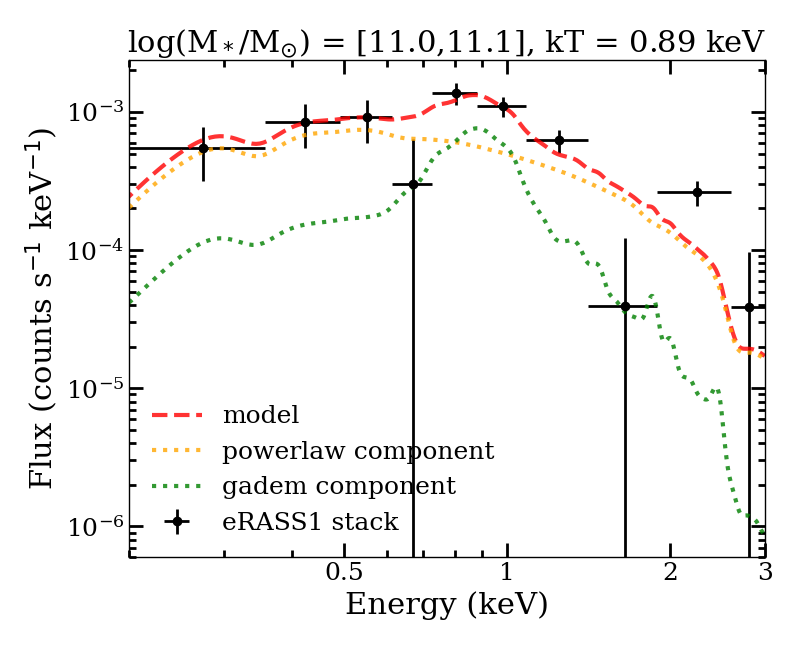}
\includegraphics[width=0.33\hsize]{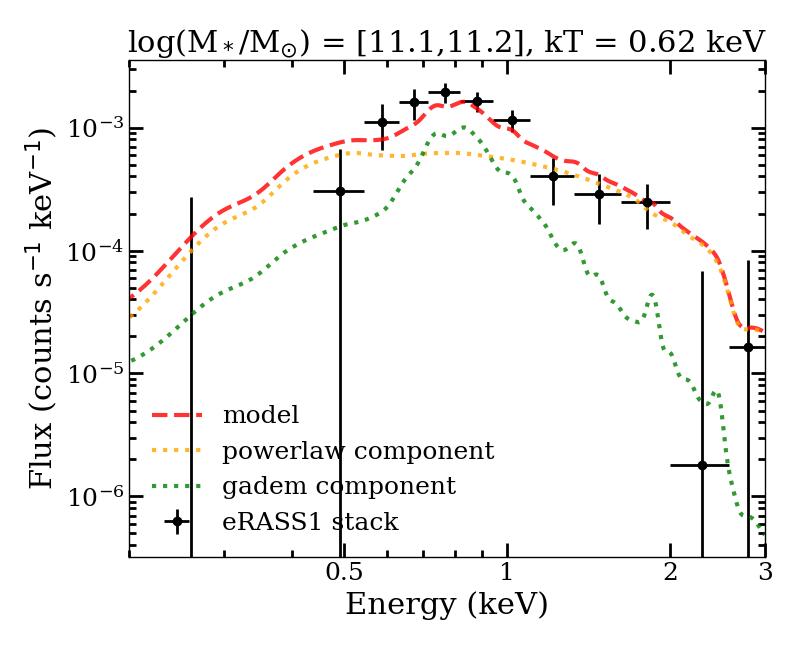}
\includegraphics[width=0.33\hsize]{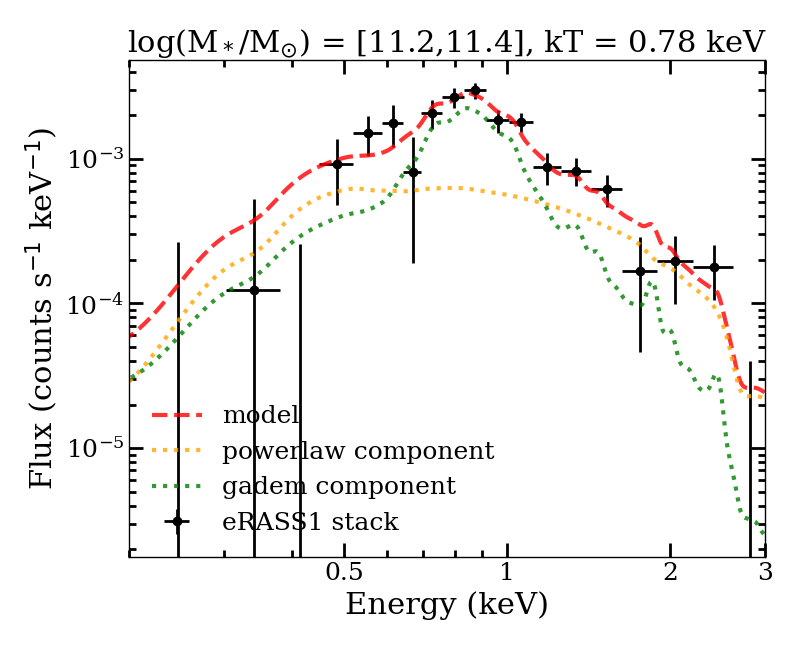}\\
\includegraphics[width=0.33\hsize]{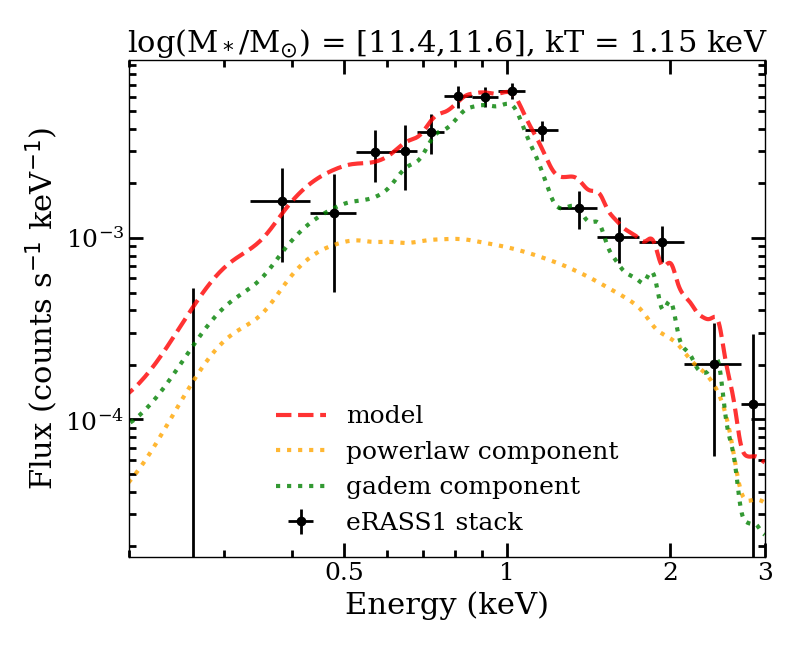}
\includegraphics[width=0.33\hsize]{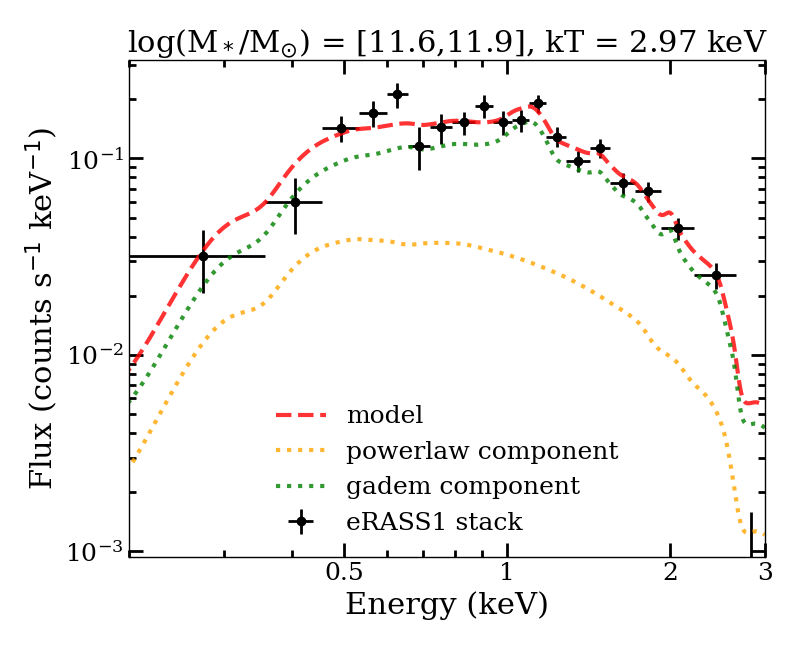}
  \caption{Stacked {background-subtracted} spectra in the 0.2--3 keV energy range for binning by BCG stellar mass. Black points show the observed stacked data, while lines represent the spectral model components: the green line is the ICM emission modeled with {\sc gadem}, the yellow line is the power-law component representing unresolved AGN, and the red line is the total model combining all components. The mean temperature of the {\sc gadem} component is provided in the title of each panel.
\label{apf:spec_mstar}}
\end{figure*}

\begin{figure*}[h!]
\centering
\includegraphics[width=0.33\hsize]{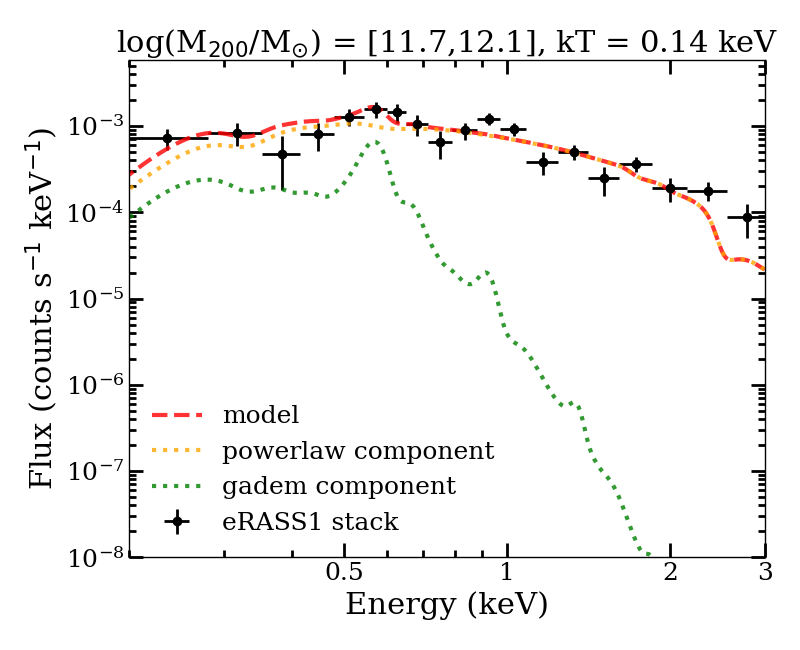}
\includegraphics[width=0.33\hsize]{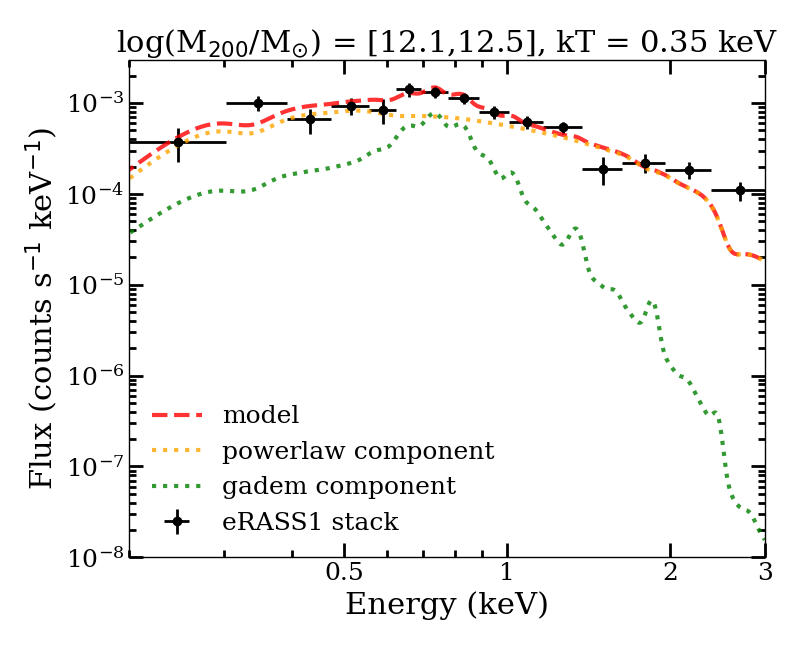}
\includegraphics[width=0.33\hsize]{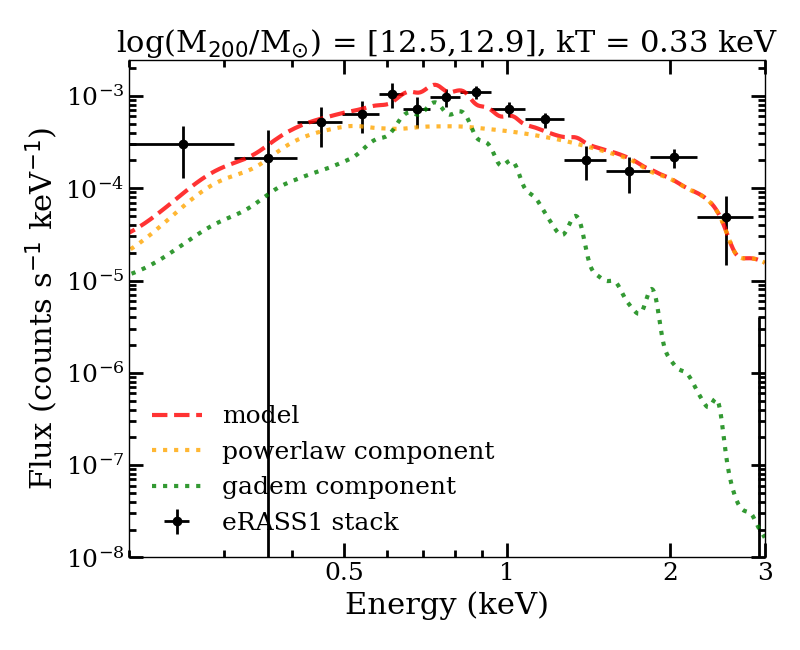}\\
\includegraphics[width=0.33\hsize]{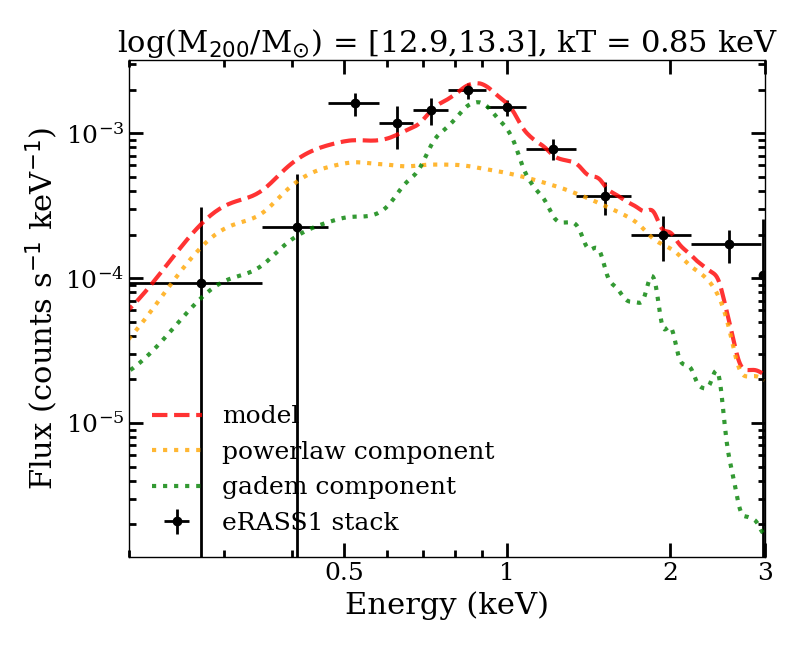}
\includegraphics[width=0.33\hsize]{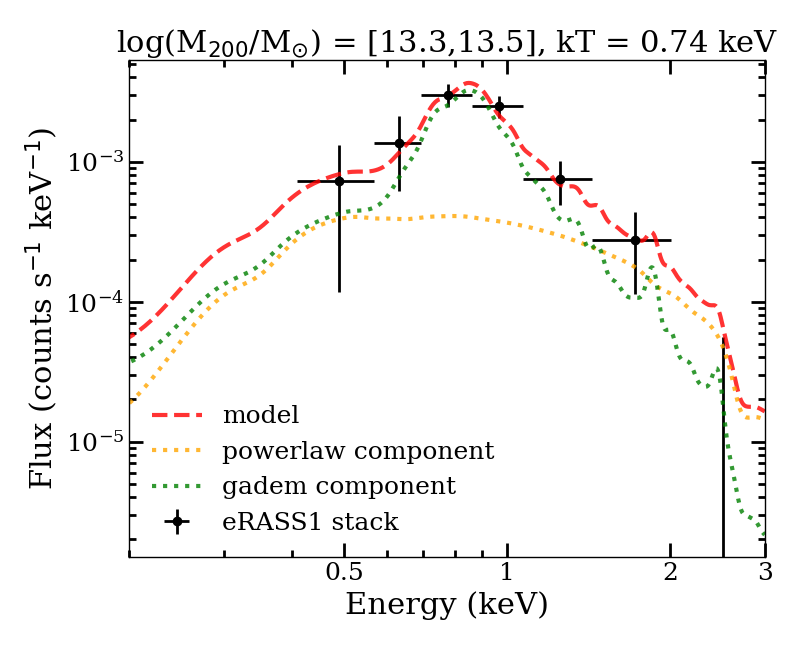}
\includegraphics[width=0.33\hsize]{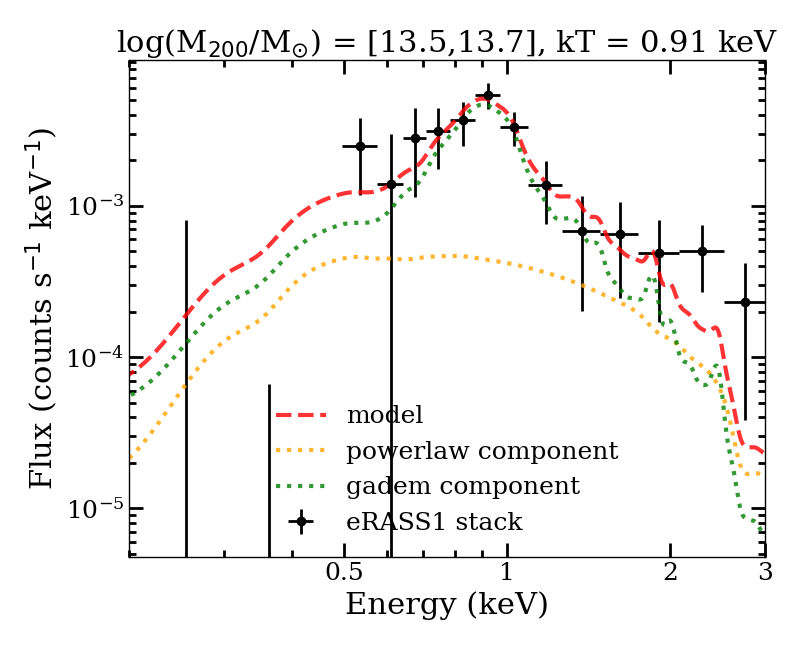}\\
\includegraphics[width=0.33\hsize]{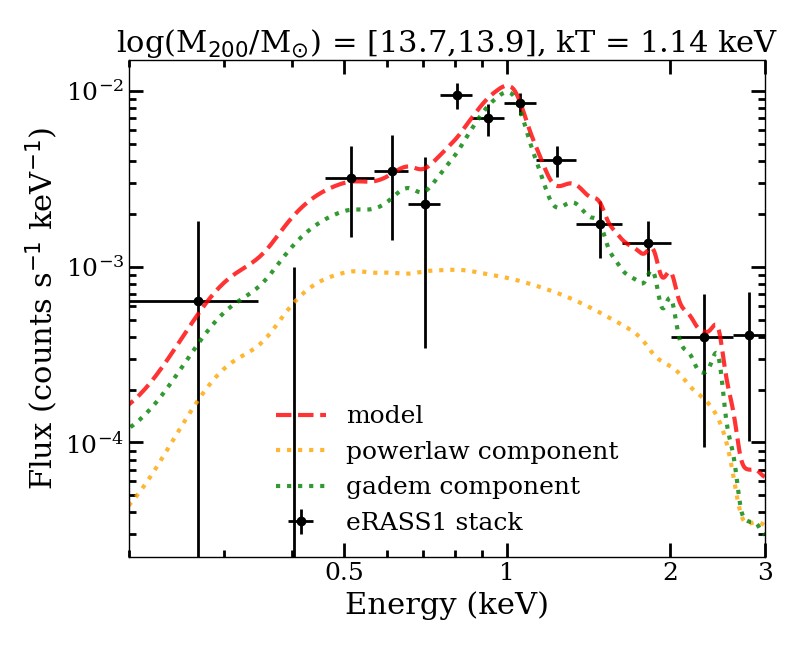}
\includegraphics[width=0.33\hsize]{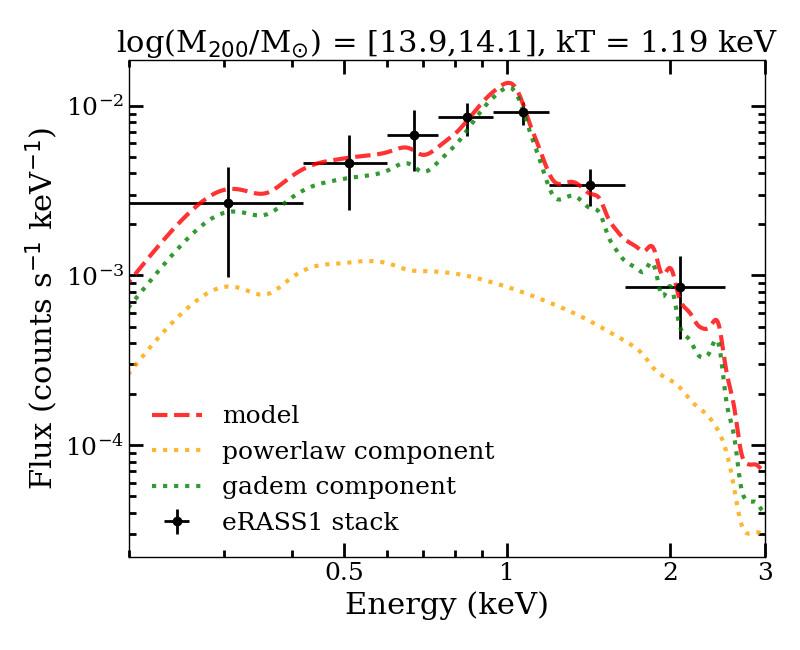}
\includegraphics[width=0.33\hsize]{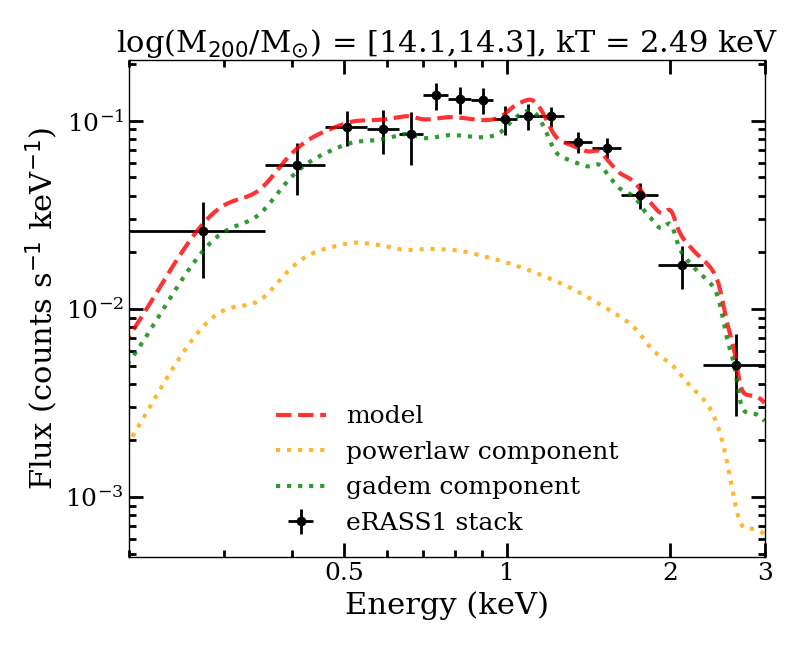}\\
\includegraphics[width=0.33\hsize]{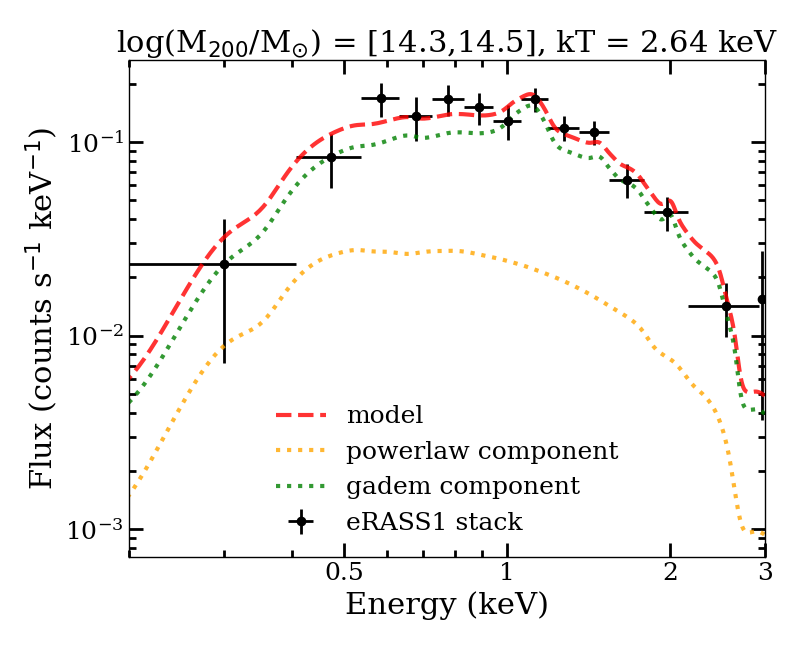}
\includegraphics[width=0.33\hsize]{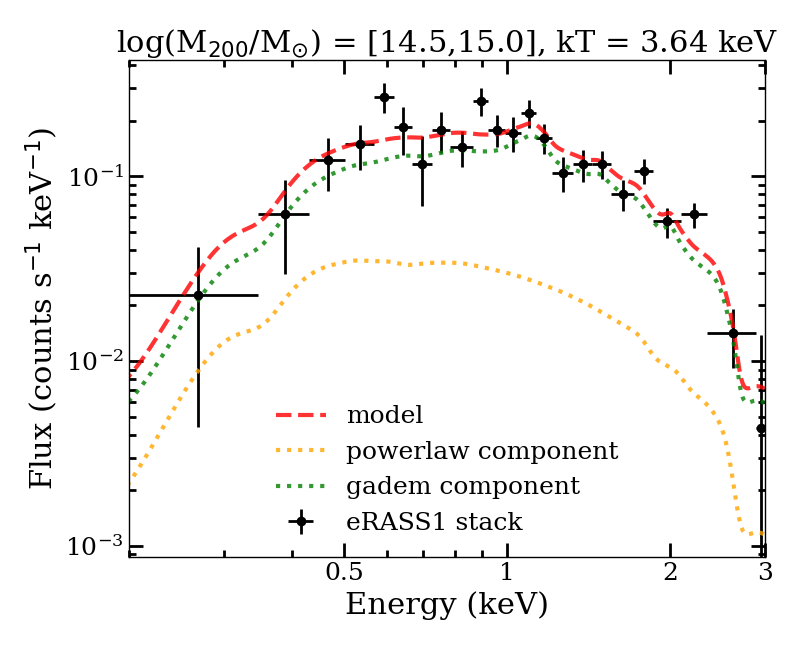}
  \caption{Same as Fig. \ref{apf:spec_mstar} but for the sample binned by halo mass based on characteristic luminosity as described in Sect. \ref{sec:sample}. \label{apf:spec_mhalo}}
\end{figure*}

In this section we present the stacked X-ray spectra in two different binning schemes {and the results of their modeling}. {Figure~\ref{apf:mh_mh} shows the comparison between the X-ray-derived halo masses and the optically based halo mass estimates. Figure~\ref{apf:tsigma} illustrates how the fit temperature depends on whether the width of the temperature distribution in the {\sc gadem} model is fixed or allowed to vary.} Figure~\ref{apf:spec_mstar} shows the stacked spectra for bins based on BCG stellar mass, while Fig.~\ref{apf:spec_mhalo} shows the stacked spectra for bins based on optically-derived halo mass from \cite{yang_galaxy_2007}. The spectral modeling results for each bin are provided in Table~\ref{tab:spectral_fit_results}.

\begin{table*}[h!]
\caption{Spectral modeling results.}   
\label{tab:spectral_fit_results}
\centering
\renewcommand{\arraystretch}{1.5}
\begin{tabular}{|c|c|c|c|c|c|c|}
\hline
log$_{10}(M_{\mathrm{bin}})$& {Bin size} & kT & dof &  {$\chi^2_\mathrm{dof}$} & \small log$_{10}(M_{200, \mathrm{X}})$ & log$_{10}(M_{\star})$\\
\small
$M_{\mathrm{bin}}$ in $M_{\odot}$ & \ & keV & \ & \ & ${M_{200,\mathrm{X}}}$ in $M_{\odot}$ & $M_{\star}$ in $M_{\odot}$
\normalsize \\
\hline
\multicolumn{7}{|c|}{Binning by $M_{200, \mathrm{optical}}$} \\
\hline
$11.7 - 12.1$ & {27344} & 0.12$^{+0.03}_{-0.01}$ & 16 & 1.88 & $11.72^{+0.24}_{-0.11}$ & $10.46^{+0.13}_{-0.14}$ \\
$12.1 - 12.5$ & {36562} & 0.36$^{+0.05}_{-0.04}$ & 12 & 2.09 & $12.65^{+0.12}_{-0.10}$ & $10.77^{+0.13}_{-0.13}$ \\
$12.5 - 12.9$ & {26431} & 0.33$^{+0.02}_{-0.01}$ & 12 & 1.45 & $12.59^{+0.05}_{-0.04}$ & $11.00^{+0.15}_{-0.12}$ \\
$12.9 - 13.3$ & {14813} & 0.85$^{+0.06}_{-0.06}$ & 9 & 1.74 & $13.37^{+0.05}_{-0.06}$ & $11.17^{+0.15}_{-0.11}$ \\
$13.3 - 13.5$ & {3718} & 0.73$^{+0.10}_{-0.09}$ & 5 & 1.00 & $13.24^{+0.09}_{-0.10}$ & $11.29^{+0.07}_{-0.06}$ \\
$13.5 - 13.7$ & {2008} & 0.91$^{+0.16}_{-0.06}$ & 13 & 1.07 & $13.43^{+0.12}_{-0.05}$ & $11.35^{+0.07}_{-0.06}$ \\
$13.7 - 13.9$ & {1092} & 1.21$^{+0.14}_{-0.16}$ & 10 & 1.40 & $13.67^{+0.07}_{-0.10}$ & $11.40^{+0.08}_{-0.06}$ \\
$13.9 - 14.1$ & {503} & 1.20$^{+0.21}_{-0.10}$ & 5 & 1.54 & $13.66^{+0.11}_{-0.06}$ & $11.48^{+0.07}_{-0.06}$ \\
$14.1 - 14.3$ & {415} & 2.49$^{+0.56}_{-0.34}$ & 13 & 1.26 & $14.27^{+0.13}_{-0.09}$ & $11.50^{+0.08}_{-0.05}$ \\
$14.3 - 14.5$ & {172} & 2.64$^{+0.81}_{-0.45}$ & 11 & 0.68 & $14.32^{+0.17}_{-0.11}$ & $11.57^{+0.07}_{-0.06}$ \\
$14.5 - 15.0$ & {84} & 3.64$^{+1.57}_{-0.80}$ & 20 & 1.22 & $14.58^{+0.22}_{-0.14}$ & $11.62^{+0.15}_{-0.10}$ \\
\hline
\multicolumn{7}{|c|}{Binning by $M_{*, \mathrm{BCG}}$} \\
\hline
$10.3 - 10.6$ & {24237} & 0.19$^{+0.03}_{-0.03}$ & 16 & 1.12 & $12.12^{+0.15}_{-0.19}$ & $10.46^{+0.09}_{-0.11}$ \\
$10.6 - 10.9$ & {11691} & 0.26$^{+0.04}_{-0.06}$ & 19 & 0.91 & $12.38^{+0.14}_{-0.23}$ & $10.75^{+0.10}_{-0.10}$ \\
$10.9 - 11.0$ & {12899} & 0.33$^{+0.20}_{-0.08}$ & 7 & 0.96 & $12.58^{+0.38}_{-0.23}$ & $10.95^{+0.03}_{-0.03}$ \\
$11.0 - 11.1$ & {13264} & 0.91$^{+0.06}_{-0.06}$ & 7 & 3.06 & $13.43^{+0.05}_{-0.05}$ & $11.05^{+0.03}_{-0.03}$ \\
$11.1 - 11.2$ & {24832} & 0.63$^{+0.05}_{-0.04}$ & 10 & 1.92 & $13.12^{+0.06}_{-0.05}$ & $11.15^{+0.03}_{-0.03}$ \\
$11.2 - 11.4$ & {17217} & 0.77$^{+0.09}_{-0.08}$ & 16 & 0.99 & $13.28^{+0.08}_{-0.08}$ & $11.29^{+0.07}_{-0.06}$ \\
$11.4 - 11.6$ & {8074} & 1.12$^{+0.09}_{-0.09}$ & 12 & 0.86 & $13.61^{+0.05}_{-0.06}$ & $11.48^{+0.08}_{-0.05}$ \\
$11.6 - 11.9$ & {984} & 2.97$^{+0.47}_{-0.31}$ & 17 & 1.36 & $14.42^{+0.09}_{-0.07}$ & $11.68^{+0.08}_{-0.05}$ \\
\hline
\end{tabular}
\tablefoot{Results of the spectral stacking and modeling for bins defined by halo mass and BCG stellar mass. For each bin, we report the mass range, {the bin size}, the mean gas temperature and its uncertainty derived from {bootstrapping of the spectral fit results}, the number of degrees of freedom (dof), the reduced $\chi^2$, the halo mass calculated using the \mbox{$M$--{$T_\mathrm{X}$}} relation from \cite{lovisari_relation}, {and the mean stellar mass in this bin from \cite{Salim}.}
}
\end{table*}

\clearpage

\section{{Validation of stacked temperatures with individual measurements}}\label{ap:individual}

\begin{figure}[h!]
\centering
\includegraphics[width=0.93\hsize]{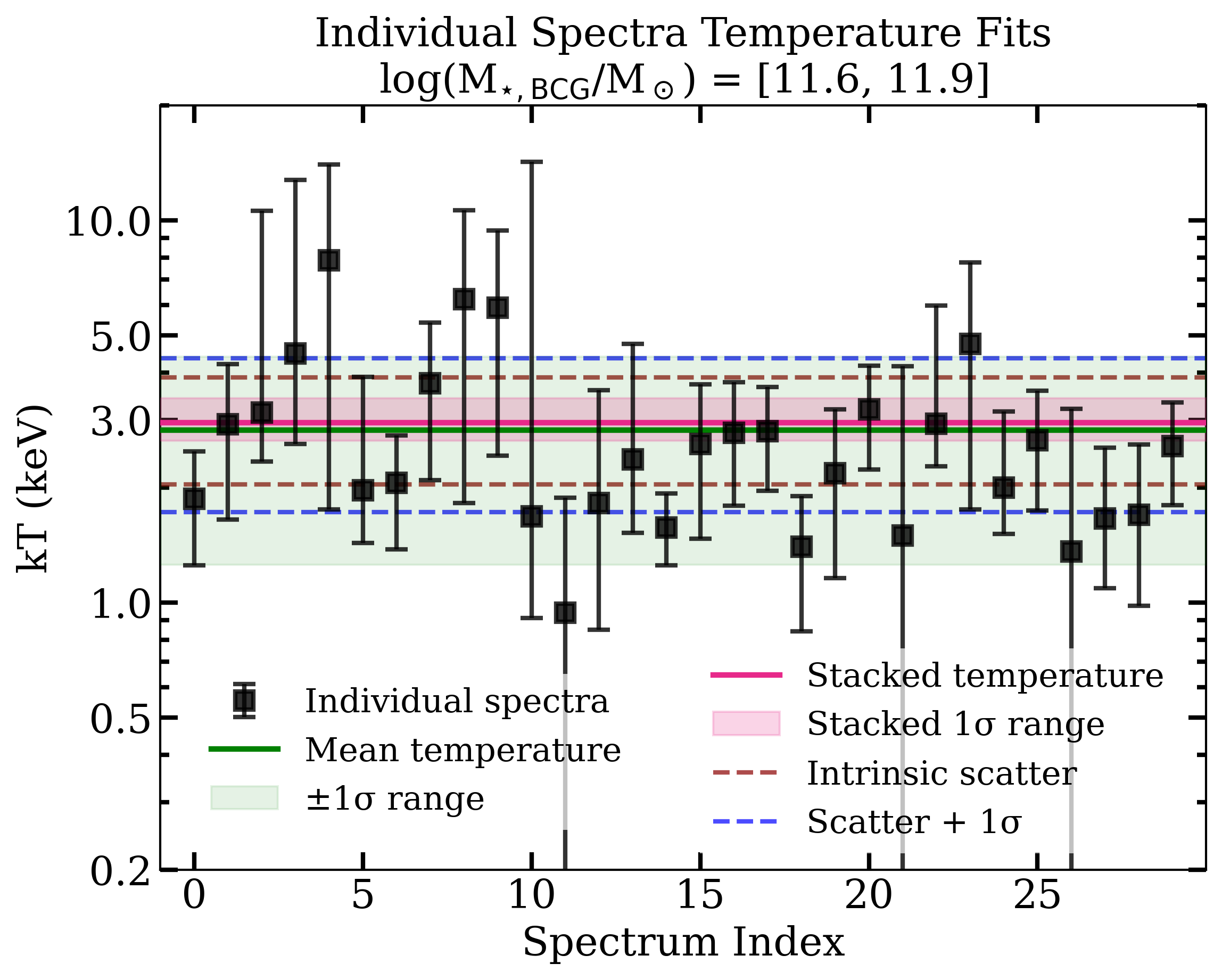}
  \caption{{Comparison of individual and stacked temperatures in the highest mass $M_{\star,\mathrm{BCG}}$ bin. Data points show the temperatures from individual spectral fits with their 1$\sigma$ errors. The solid pink line and shaded region show the stacked temperature and its 1$\sigma$ uncertainty from Table~\ref{tab:spectral_fit_results}. Dashed red lines indicate the model width of the intrinsic temperature distribution $\sigma_{T}$. The solid green line and shaded region correspond to the mean temperature of the individual measurements with its 1$\sigma$ uncertainty.}}
  \label{apf:individual}
\end{figure}

{As a consistency check, we compare the stacked temperature in the highest mass $M_{\star,\mathrm{BCG}}$ bin with the temperatures obtained from individual spectral fits, since in this bin all the sources are bright enough for individual detection. Figure~\ref{apf:individual} shows the individual measurements, the stacked result with its 1$\sigma$ statistical uncertainty, the model width of the temperature distribution $\sigma_{T}$ (for more details, see Sect.~\ref{sec:fitting} and Fig.~\ref{apf:tsigma}), and the mean of the individual measurements with its 1$\sigma$ uncertainty. This comparison demonstrates that the mean of the individual measurements closely matches the stacked temperature, and that the combination of stacking uncertainty and the intrinsic temperature distribution width reproduces the scatter of individual systems, confirming the reliability of the method.}

\end{appendix}

\end{document}